\documentclass[preprint]{aastex62}
\usepackage{amsmath,amssymb,CJK,soul}
\usepackage{longtable}
\usepackage{hyperref}
\usepackage{longtable}
\usepackage{threeparttablex} 
\usepackage{textcomp}
\usepackage{lineno}

\newcommand{\yr}{{2024 YR$_4$ }}
\newcommand{\yrns}{{2024 YR$_4$}}

\received{--}
\revised{--}
\accepted{--}
\submitjournal{ApJL}

\shorttitle{Discovery and characterization of \yrns}
\shortauthors{Bolin et al.}

\begin{document}

\title{The discovery and characterization of Earth-crossing asteroid \yr}

\author[0000-0002-4950-6323]{Bryce T. Bolin}
\altaffiliation{These authors contributed equally to this work.}
\affiliation{Eureka Scientific, Oakland, CA 94602, U.S.A.}

\author[0000-0002-2934-3723]{Josef Hanu\v{s}}
\altaffiliation{These authors contributed equally to this work.}
\affiliation{Charles University, Faculty of Mathematics and Physics, Institute of Astronomy, 180 00 Prague, Czech Republic}

\author[0000-0002-7034-148X]{Larry Denneau}
\altaffiliation{These authors contributed equally to this work.}
\affiliation{Institute for Astronomy, University of Hawai`i at M\={a}noa, Honolulu, HI, 96822}

\author[0009-0001-6475-144X]{Roberto Bonamico}
\affiliation{ BSA Osservatorio (K76), 12038 Savigliano, Cuneo, Italy}

\author[0009-0000-8709-5273]{Laura-May Abron}
\affiliation{Griffith Observatory, Los Angeles, CA 90027}

\author[0000-0002-8963-2404]{Marco Delbo}
\affiliation{Universit\'{e} C\^{o}te d'Azur, CNRS Lagrange, Observatoire de la C\^{o}te d'Azur, F 06304, Nice, France}
\affiliation{University of Leicester, School of Physics and Astronomy, Leicester, LE1 7RH, UK}

\author[0000-0003-4914-3646]{Josef \v{D}urech}
\affiliation{Charles University, Faculty of Mathematics and Physics, Institute of Astronomy, 180 00 Prague, Czech Republic}

\author[0000-0001-7830-028X]{Robert Jedicke}
\affiliation{Institute for Astronomy, University of Hawai`i at M\={a}noa, Honolulu, HI, 96822}

\author[0000-0002-2250-8687]{Leo Y. Alcorn}
\affil{W. M. Keck Observatory, Waimea, HI 96743, USA}

\author[0000-0001-7101-9831]{Aleksandar Cikota}
\affil{International Gemini Observatory/NSF NOIRLab, Casilla 603, La Serena, Chile}

\author[0000-0002-5854-7426]{Swayamtrupta Panda}
\affil{International Gemini Observatory/NSF NOIRLab, Casilla 603, La Serena, Chile}

\author[0000-0001-6533-6179]{Henrique Reggiani}
\affil{International Gemini Observatory/NSF NOIRLab, Casilla 603, La Serena, Chile}




\begin{abstract}
We describe observations and physical characteristics of Earth-crossing asteroid \yrns, discovered on 2024 December 27 by the Asteroid Terrestrial-impact Last Alert System. The asteroid has semi-major axis, $a$ = 2.52 au, eccentricity, $e$ = 0.66, inclination $i$ = 3.41$^{\circ}$, and a $\sim$0.003 au Earth minimum orbit intersection distance. We obtained g, r, i, and Z imaging with the Gemini South/Gemini Multi-Object Spectrograph on 2025 February 7 and Y and J imaging with the Keck/Multi-Object Spectrometer For Infra-Red Exploration on 2025 February 12. We measured a g-i spectral slope of 13$\pm$3 $\%$/100 nm, and color indices g-r = 0.70 $\pm$ 0.10, r-i = 0.25$\pm$0.06, i-Z = -0.27 $\pm$ 0.10, and Y-J = 0.41 $\pm$ 0.10. \yr has a spectrum that best matches R-type and Sa-type asteroids and a diameter of $\sim$30-65 m using our measured absolute magnitude of 23.9 $\pm$ 0.3 mag, and assuming an albedo of 0.15-0.4. The lightcurve of \yr shows $\sim$0.4 mag variations with a rotation period of $\sim$1170 s. We use photometry of \yr from Gemini and other sources taken between 2024 December to 2025 February to determine the asteroid's spin vector and shape, finding that it has an oblate, $\sim$3:1 a:c axial ratio and a pole direction of $\lambda$, $\beta$ =  $\sim$42$^{\circ}$, $\sim$-25$^{\circ}$. Finally, we compare the orbital elements of \yr with the NEO population model and find that its most likely sources are resonances between the inner and central Main Belt.
\end{abstract}

\section{Introduction}
Earth-crossing asteroids (ECAs) are asteroids whose orbital trajectories cross the Earth's coming within its 0.983 au perihelion, $q$, and 1.017 au aphelion, $Q$, distances. By definition, ECAs belong to the Apollo NEO class with semi-major axes, $a$$>$1 au, and $q$ $<$ 1.017 au, and the Aten NEO class with $a$ $<$1 au and $Q$ $>$ 0.983 au. ECAs present scientific opportunities for detailed study of close approaching near-Earth objects \citep[NEO, e.g.,][]{Granvik2013,Binzel2015}. Interactions with the Earth's tides during close-approaching asteroids can affect the structural integrity of asteroids, causing changes in their surface material properties \citep[][]{Binzel2010}, or completely disrupt, making Earth-encounters a source of some NEOs \citep[][]{Granvik2024,Nesvorny2024NEOMOD3}. Additionally, a portion of ECAs are a source of impacts on the Earth and the Moon affecting their total impact flux \citep[][]{Mazrouei2019,Cohen2023}. Thus, ECAs provide examples of useful study on the effect of close approaches on the physical properties of asteroids and the impact chronology of the Earth and Moon.

The majority of ECAs originate from sources in the inner Main Belt between 2.2 au and 2.5 au \citep[][]{Granvik2018}. Asteroids leave the Main Belt due to the Yarkovsky effect causing their semi-major axes to evolve, increasing for prograde-spinning asteroids and decreasing for asteroids that are spinning retrograde \citep[][]{Vokrouhlicky2015}. As asteroids evolve in their semi-major axis due to the Yarkosvky effect, they cross resonances such as the $\nu_{6}$ resonance that defines the inner border of the Main Belt at 2.2 au, the 3:1 mean motion resonance (MMR) with Jupiter at 2.5 au, and various weaker resonances located between 2.2 au and 2.5 au \citep[][]{Granvik2017}. The interaction between the orbits of these asteroids and the resonances causes their orbits to chaotically evolve, sending them on trajectories where they can encounter the Earth \citep[][]{Wisdom1983,Farinella1993}. 

While the majority of inner Main Belt asteroids belong to the siliceous S-complex, $\sim$20$\%$ of inner Main Belt asteroids belong to the more primitive and carbonaceous C-complex \cite[][]{DeMeo2013aa}. This results in the majority of C-complex NEOs originating from the inner region of the Main Belt \citep[][]{Binzel2019}. Additionally, a considerable portion of NEOs can also belong to the X-complex, whose compositions span enstatite chondrites and achondrite, metallic and stony iron compositions \citep[][]{Cloutis1990, Bus2002, Vernazza2009, DeMeo2015, Avdellidou2022}. While the orbits of NEOs and their likely sources within the Main Belt can provide constraints on their composition \citep[][]{Jedicke2022}, the information on the composition of NEOs is determined directly through observations covering visible and near-infrared wavelengths \citep[][]{DeMeo2009}.

This paper discusses the discovery of ECA \yr by the Asteroid Terrestrial-impact Last Alert System \citep[ATLAS,][]{Tonry2018} telescope and observations taken at Gemini South and the determination of its physical shape model. We also adapt the techniques described by \citep[][]{Bolin2020CD3,Bolin2021LD2,Bolin2025PT5} of using multi-band photometry to constrain the physical and taxonomic properties of \yrns. Finally, we will use the lightcurve inversion techniques established by \citep[][]{Hanus2011,Hanus2018} on publically sparse and dense data to determine the spin vector and shape of \yrns, and test its likely source within the Main Belt using the NEO population model \citep[][]{Morbidelli2020albedo,Nesvorny2023NEO,Nesvorny2024NEOMOD3}.

\section{Observations}
\label{s.obs}

The ATLAS Chile, Rio Hurtado telescope, with Minor planet Center (MPC) observing code W68, made the first discovery observations of \yr on 2024 December 27 05:42:49 UTC as seen in Panels a-d of Fig~1. The asteroid was discovered at right ascension (RA) and declination (dec) of 08:56:40.97, -00:16:11.93, and was located 0.996 au from the Sun, 0.017 au from the Earth, and at a phase angle of 43.2$^{\circ}$, and an apparent magnitude of $\sim$16.17 in the ATLAS o-band filter \citep[][]{Denneau2024YR4}. The ATLAS camera has a pixel scale of 1.86\arcsec, an ``orange'' o-band custom filter providing wavelength coverage between 560 nm and 820 nm, and an effective wavelength of 663 nm \citep[][]{Tonry2018}. The asteroid was detected in all four o-band images taken in a 3160 s interval from 2024 December 27 05:42:49 UTC to 2024 December 27 06:35:30 UTC. The asteroid was moving northwest at $\sim$22\arcsec/min, resulting in the detections forming a trail $\sim$11\arcsec~long (panels a-d, Fig.~1).

\begin{figure}
\centering
\includegraphics[scale=0.5]{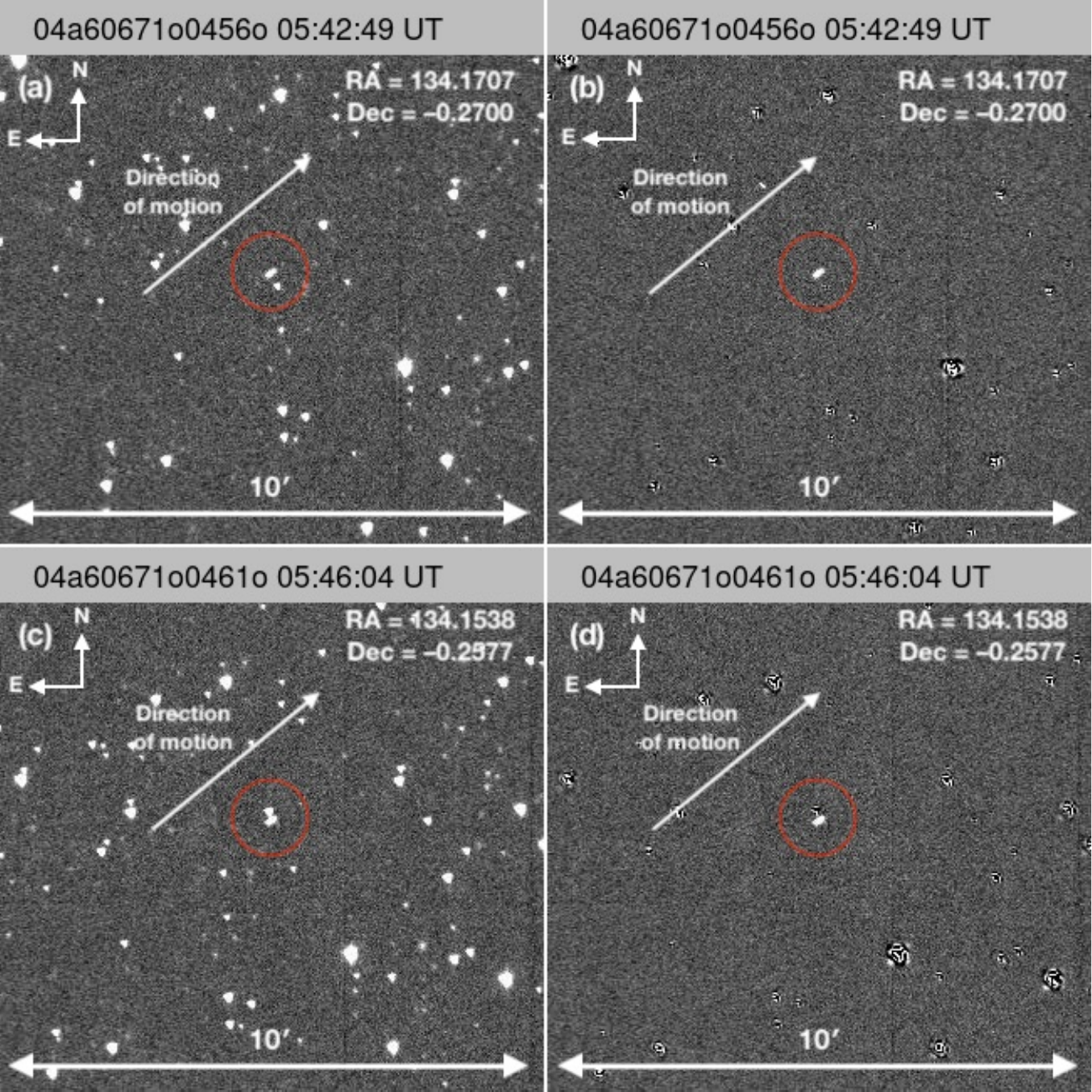}
\caption{\textbf{Panel a:} The first of four o-band ATLAS-Chile telescope discovery images of \yr
from 2024 December 27 05:42:49 UTC. The asteroid moved at a rate of 22.8 arcminutes per hour (9.11 degrees
per day) in the northwest direction. The asteroid makes a $\sim$6 pixel trail ($\sim$11~\arcsec) in the 30 s ATLAS exposures, indicated by the red circle. \textbf{Panel b:} the same as panel a but shows the image after subtracting static sources. Despite the presence of nearby stars, the asteroid is detected cleanly in the subtracted images. \textbf{Panel c:} the same as panel a, but the second of four o-band images containing \yr taken on 2024 December 27 05:46:04 UTC. \textbf{Panel d:} the same as panel c but shows the image after subtracting static sources. The asteroid was detected with an apparent magnitude of o=16.54 in panel a and o=15.80 in panel c. The large black areas in the subtracted images are regions of saturated pixels from bright stars. The direction of the asteroid and cardinal directions are indicated in each figure.}
\end{figure}

A director's discretionary time request (DDT) was granted at Gemini South to observe \yr on 2025 February 7 02:23:46 UTC under program GS-2025A-DD-102 (PI: Bolin). The asteroid was located at RA, dec = 08:04:08.66, +13:33:24.31, 1.367 au from the Sun, 0.399 au from the Earth and had a phase angle of 14.8$^{\circ}$ during the 2025 February 7 UTC observations. The Gemini Multi-Object Spectrograph (GMOS) instrument \citep[][]{Hook2004} was used in imaging mode to observe \yr in g, r, I, and Z filters. The GMOS camera comprises a 5.5\arcmin~$\times$ 5.5\arcmin~ field of view Hamamatsu array with an effective pixel scale of 0.08\arcsec. The camera was used in 2 $\times$ 2 binning mode, giving an effective pixel scale of 0.16\arcsec. The GMOS g, r, and i filers are equivalent to Sloan Digital Sky Survey (SDSS) g, r, and i filters with effective wavelengths of 475 nm, 630 nm, and 780 nm \citep[][]{Fukugita1996}. 

A filter was used for coverage past 800 nm equivalent to the the WFCAM Z with an effective wavelength of 880 nm \citep[][]{Casali2007}. This filter has the advantage of being equivalent to the SDSS z filter, but avoids telluric absorption features near 950 nm \citep[][]{Hodgkin2009}. Observations of \yr were also taken on 2025 February 23 02:13:17 UTC when the asteroid was located at RA, dec = 08:04:08.66, 13:33:24.31, 1.520 au from the Sun, 0.598 au from the Earth, and had a phase angle of 21.7$^{\circ}$, but only exposures in i-band were taken. During both the 2025 February 7 UTC and 2025 February 23 UTC observing dates, 200 s exposures were used for all filters, and the telescope was tracked at the rate of motion of \yrns.

During the 2025 February 7 UTC Gemini S observations, the measured seeing in the r band was $\sim$0.8\arcsec\, and the airmass ranged from 1.43 -1.50 during the $\sim$2 h 35 min observing sequence. We acquired 11 g band images, 12 r band images, 8 i band images, and 8 Z band images. The g, r, i, and Z filters changed between exposures in a rgiZrgriZgr sequence to smear out the brightness variations in the asteroid's lightcurve and limit its effect on the color measurements. The \texttt{DRAGONS} image pipeline was used with bias and flat field calibration frames to detrend the data \citep[][]{Labrie2023}. Images of \yr when it intersected a background star were discarded. A mosaic showing \yr in the co-added g, r, i, and Z frames is shown in Fig~2.

Additionally, we obtained observations of \yr with Keck I on 2025 February 12 07:26:30 UTC using the Multi-Object Spectrometer For Infra-Red Exploration (MOSFIRE) under program R379 (PI: Bolin). On 2025 February 12 UTC, \yr had a heliocentric distance of 1.417 au, a geocentric distance of 0.461 au, and a phase angle of 17.4$^{\circ}$. MOSFIRE has a plate scale of 0.18\arcsec/pixel and was used in imaging mode \citep[][]{McLean2012}. Observations of \yr were taken with the Y filter (possessing a central wavelength of 1048 nm and FWHM of152 nm)  and the J filter (possessing a central wavelength of 1253 nm and of FWHM 200 nm). Seeing measured for stars in fields near the asteroid in J band was $\sim$0.8\arcsec. The telescope was tracked at the non-sidereal motion of \yr, and an exposure time of 180 s was used. A nearby solar analog star was also observed to provide a calibration reference.

\begin{figure}
\centering
\includegraphics[scale=.27]{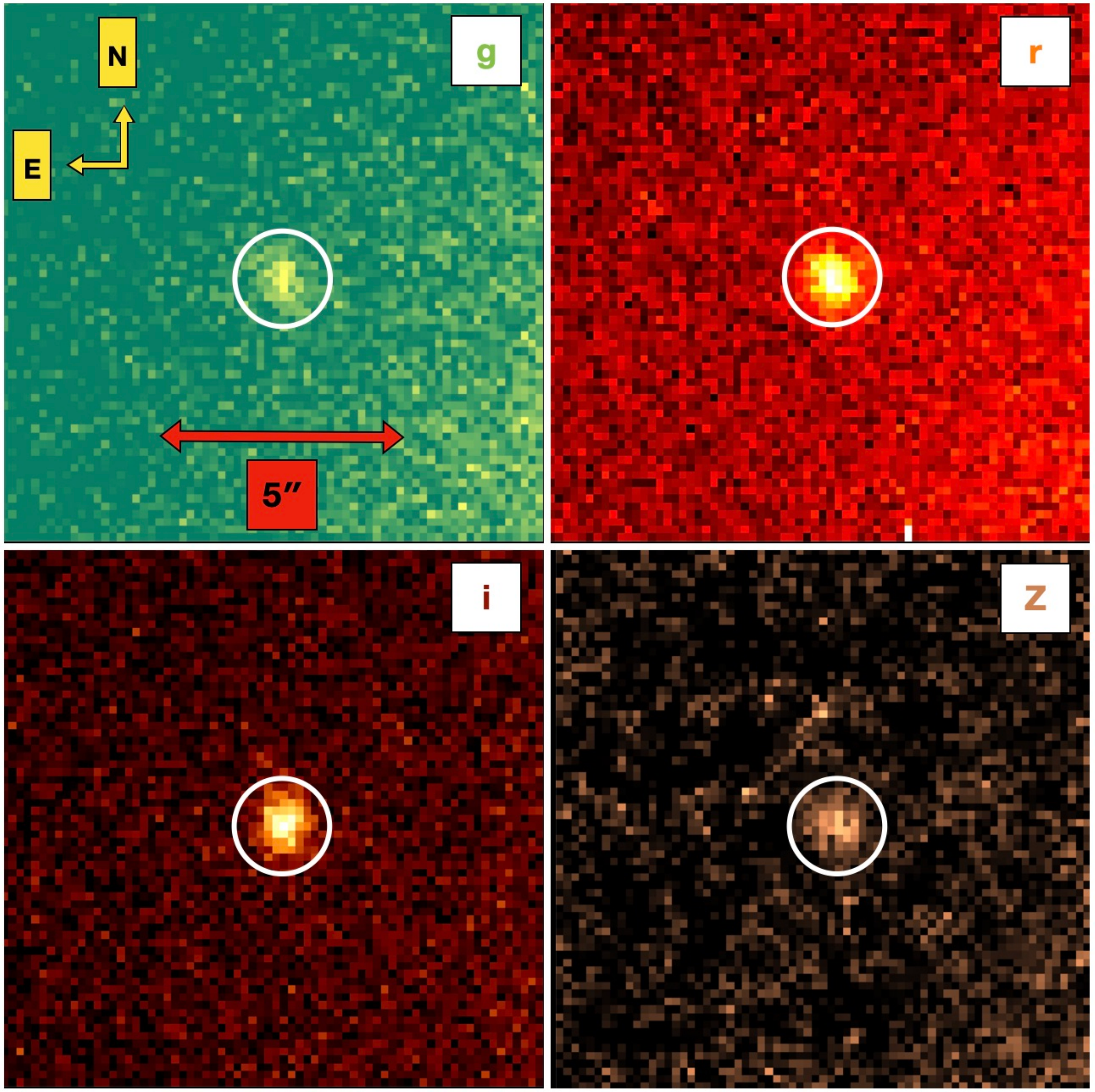}
\caption{\textbf{Top left panel:} a co-add stack of 11 x 200 s g filter images of \yr with a red 5\arcsec~wide double arrow for scale. The cardinal directions are shown with yellow arrows. \textbf{Top right panel:} a co-add stack of 12 x 200 s r filter images of \yrns. \textbf{Bottom left panel:} a co-add stack of 8 x 200 s i filter images of \yrns. \textbf{Bottom right panel:} a co-add stack of 8 x 60 s Z filter images of \yrns. The scale and orientation of the r, i, and Z band panels are the same as in the g band panel.}
\end{figure}

\section{Results}
\subsection{Orbit and source population}

Astrometry of \yr from observing dates was reported to the MPC \citep[][]{Williams2025OrbitUpdate,Williams2025MPS}. Additional astrometry was reported by $\sim$440 observations of \yr between 2024 December 25 UTC and 2025 March 1 UTC \footnote{\url{https://minorplanetcenter.net/db_search/show_object?utf8=\%E2\%9C\%93&object_id=2024+YR4}, accessed on 2025 March 4 UTC.}. Table~1 shows the six orbital element solution computed by JPL HORIZONS using these observations\footnote{\url{https://ssd.jpl.nasa.gov/tools/sbdb_lookup.html\#/?sstr=2024\%20YR4}, solution date 2025 March 1 UTC, accessed on 2025 March 4 UTC.}. The asteroid has an Apollo-type orbit with $a$$>$1 au, and a perihelion takes it inside the orbit of the Earth. The Earth minimum orbit intersection distance (MOID) according to JPL HORIZONS is 2.82$\times10^{-3}$ au.

\begin{table}
\centering
\caption{Orbital elements of \yr taken from JPL HORIZONS (accessed on 2025 March 4 UTC) based on $\sim$440 observations submitted to the MPC between 2024 December 25 UTC and 2025 March 1 UTC. The osculating orbital elements are shown for the Julian date (JD) epoch 2,460,800.5. The 1~$\sigma$ uncertainties are given in parentheses.}
\label{t:orbit}
\begin{tabular}{ll}
\hline
Heliocentric Elements&
\\ \hline
Epoch (JD) & 2,460,800.5\\
\hline
Time of perihelion, $T_p$ (JD) & 2,460,636.9175586$\pm$(3.76x10$^{-5}$)\\
Semi-major axis, $a$ (au) & 2.5158656$\pm$(2.35x10$^{-5}$)\\
Eccentricity, $e$ & 0.66154790$\pm$(3.35x10$^{-6}$)\\
Perihelion, $q$ (au) & 0.851499970$\pm$(4.87x10$^{-7}$)\\
Aphelion, $Q$ (au) & 4.1802311$\pm$(3.90x10$^{-5}$)\\
Inclination, $i$ ($^{\circ}$) & 3.4081731$\pm$(1.22x10$^{-5}$)\\
Ascending node, $\Omega$ ($^{\circ}$) & 271.36561725$\pm$(8.89x10$^{-6}$)\\
Argument of perihelion, $\omega$ ($^{\circ}$) & 134.3613663$\pm$(1.54x10$^{-5}$)\\
Mean Anomaly, $M$ ($^{\circ}$) & 40.402628$\pm$(5.7x10$^{-4}$)\\
\hline
\end{tabular}
\end{table}

The orbital and location of \yr at the time of its discovery on 2024 December 27 UTC is shown on Fig.~3 based on values from JPL Horizons\footnote{\url{https://ssd.jpl.nasa.gov/horizons/app.html\#/}}. We compared the orbit elements, $a$, $e$, $i$, and absolute magnitude $H$ (see following Section 3.2) of \yr with the NEOMOD3 population model \citep[][]{Nesvorny2024NEOMOD3}, suggesting its most likely source is the 3:1 MMR with Jupiter resonance located in between the inner and central Main Belt regions at 2.5 au with a 73.9$\%$ probability. The next most likely location, with a 9.8$\%$ chance is the $\nu_{6}$ resonance located at the inner edge of the Main Belt at 2.2 au. We combine our source probabilities for \yr with the NEO albedo model of \citet[][]{Morbidelli2020albedo} to estimate its albedo, $p_v$, finding a predicted value of $\sim$0.18.

\begin{figure}
\centering
\includegraphics[scale=0.25]{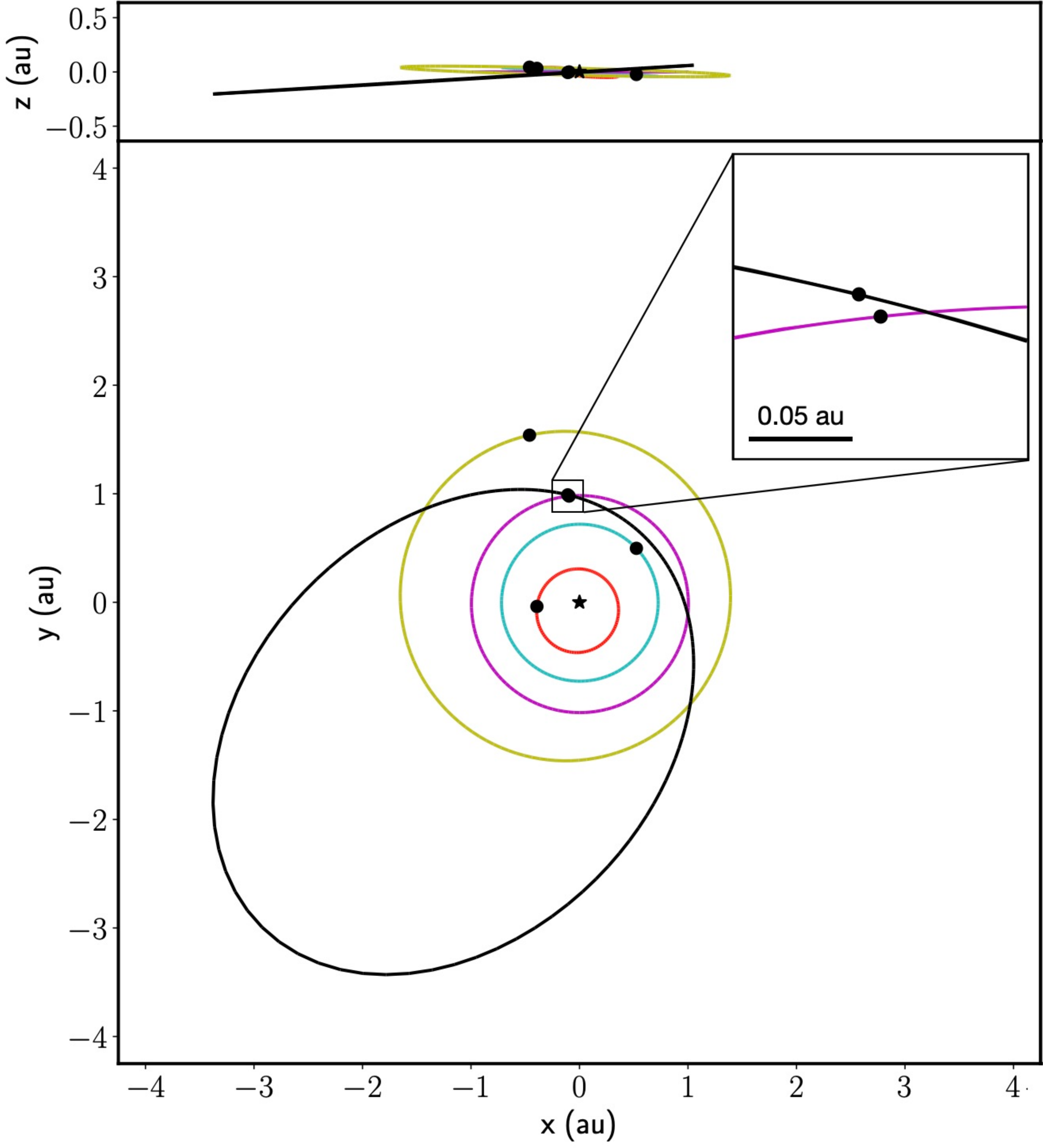}
\caption{\textbf{Top panel:} Edge-on ecliptic view of the orbits and positions of \yr (black), Mercury (red), Venus (blue), Earth (purple), and Mars (green) at the time of the discovery of \yr on 2024 December 27 05:42:49. \textbf{Bottom panel:} the same as the top panel, except from the ecliptic pole. A close-up of the positions of \yr and the Earth is shown in an insert.}
\end{figure}

\subsection{Color photometry and spectral classification}
\label{sec:photo}

We measured the photometry of \yr in the Gemini g, r, i, and Z images using a 1.12\arcsec~aperture and a sky-median subtraction annulus with and inner and outer radius of 2.08\arcsec~and 2.88\arcsec. Local solar analog stars from the Pan-STARRS catalog \citep[][]{Chambers2016} were used to calibrate the photometry according to the transformations described by \citet[][]{Tonry2012}. The brightnesses per bandpass of our measured photometry are g = 24.01$\pm$0.09, r = 23.31$\pm$0.04, i = 23.06$\pm$0.04, Z = 23.34$\pm$0.09. The asteroid's g-i color and spectral gradient are g-i = 0.95$\pm$0.10 and 13$\pm3$3$\%$/100 nm. The other multi-band color indices of \yr are g-r = 0.70$\pm$0.10, r-i = 0.25$\pm$0.06, i-Z = -0.27$\pm$0.10, r-Z = -0.02$\pm$0.10, and the a$^{*}$ parameter, an indicator of spectral slope defined by \citet[][]{Ivezic2001} as a$^{*}$ = (0.89 (g-r)) + (0.45 (r-i)) - 0.57 is a$^{*}$ = 0.16$\pm$0.09. We also measured the 2025 February 12 UTC Keck/MOSFIRE J and K photometry of both \yr and the solar analog using a 0.9\arcsec~aperture and an inner and outer radius of 1.44\arcsec~and 2.16\arcsec. The per-bandpass brightness of our measured photometry is Y = 23.56 $\pm$ 0.07, and J = 23.15 $\pm$ 0.07 resulting in a color index of Y-J = 0.41 $\pm$ 0.10.

The a$^{*}$ and i-z(Z) of \yr are plotted in Fig.~4, which has a$^{*}$ and i-z(Z) on the border between the S-complex asteroids, which have on average a$^{*}$ = 0.12$\pm$0.03 and i-z = -0.08 $\pm$ 0.07, and the V-complex asteroids, which have on average a$^{*}$ = 0.15 $\pm$ 0.11 and i-z = -0.46 $\pm$ 0.04 \citep[][]{DeMeo2009}. However, this is still redder compared to the Sun which has a$^{*}$ = -0.13 and i-z = 0.03 \citep[][]{Willmer2018}, and C-complex asteroids, which have on average a$^{*}$ = -0.09$\pm$0.02 and i-z = 0.02$\pm$0.03 \citep[][]{DeMeo2009}. Additionally, the equivalent V band brightness of \yr is V = 23.92$\pm$0.10 determined by combining our g and r band measurements of the asteroid and the color transformation from \citep[][]{Jordi2006}. The  Y-J color of 0.41 $\pm$ 0.10 measured from our MOSFIRE photometry is broadly consistent with the Y-J colors of S-complex asteroids \citep[][]{Popescu2018MOVIS}.

\begin{figure}
\centering
\includegraphics[scale=.405]{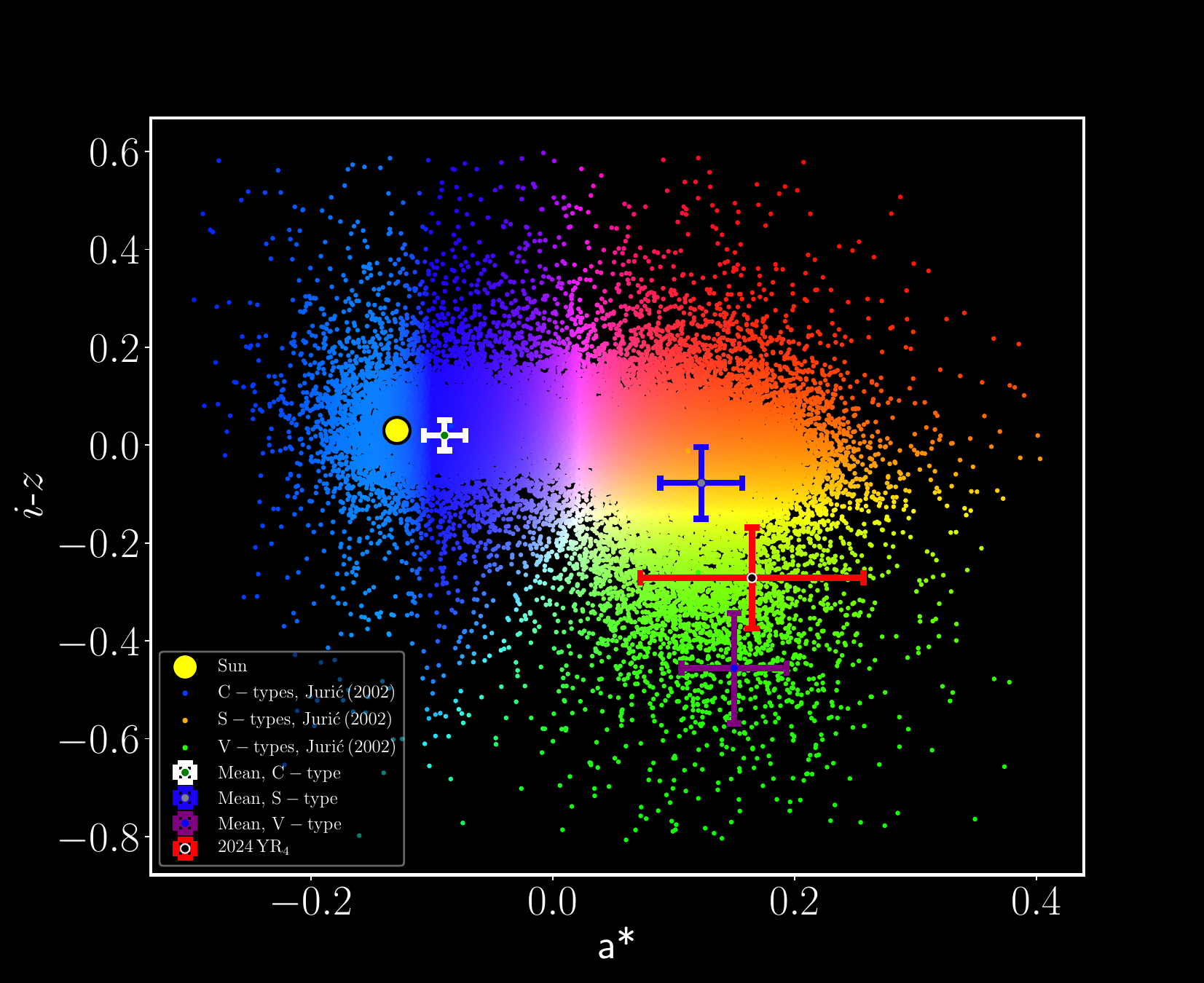}
\caption{The a$^{*}$ parameter vs. i-z(Z) colors of \yr from our measurements, and the a$^{*}$ and i-z(Z) of
 C-, S-, and V-complex asteroids from \citet{Ivezic2001,Juric2002}. We use the scheme of \citet[][]{Ivezic2002} that sets the color of data points as a function of a$^{*}$ and i-z where blue symbol colors correspond to C-complex asteroids, red symbol colors correspond to S-complex asteroids and green symbol colors correspond to V-complex asteroids. We note we use the Z magnitude of \yr in place of its Z magnitude for its $i$-$z$ color calculation used in this plot. We also include the a$^{*}$ and $i$-$z$ determined with the mean S, V, and C-complex spectra taken from \citet[][]{DeMeo2009}, and the a$^{*}$ and $i$-$z$ determined from the Sun's colors \citep[][]{Willmer2018}.}
\end{figure}

We estimate the H magnitude of \yr using our estimated $V$ magnitude and the HG phase function \citep[][]{Bowell1988}:
\begin{equation}
\label{eqn.brightness}
H = V - 5\, \mathrm{log_{10}}(r_h \Delta) +2.5\,\mathrm{log_{10}}\left[ (1 - G)\,\Phi_1(\alpha) + G\,\Phi_2(\alpha) \right ]
\end{equation}
where $r_h$ is the 1.367 au heliocentric distance of \yrns, $\Delta$ is its geocentric distance of 0.399 au and $\alpha$ is its phase angle of 14.8$^\circ$ of \yr on 2025 February 7 UTC. $G$ is the phase coefficient which we use the value of 0.23, the average value for S-complex asteroids \citep[][]{Pravec2012,Veres2015}. $\Phi_1(\alpha)$ and $\Phi_2(\alpha)$ are the basis functions normalized at $\alpha$ = 0$^\circ$ described in \citep[][]{Bowell1988}. We obtain $H$ = 23.91$\pm$0.10 in agreement from the value calculated from JPL HORIZONS of $H$ = 23.96$\pm$0.28 but caution that the uncertainty on estimate of $H$ is underestimated due to the lack of information about the phase function \citep[e.g.,][]{Bolin2022IVO}. Therefore, we adopt the larger uncertainty on $H$ taken from JPL HORIZONS of $\sim$0.3.

For each g, r, i, and Z filter, we compute the spectral reflectance of \yrns, dividing its measured flux by the flux of a solar analog star. The reflectivity spectrum of \yr normalized to a wavelength of 550 nm is shown in Fig.~5, and features a red slope from 470 nm to 770 nm, and absorption at 880 nm. We take the approach of \citet[][]{Bolin2023Dink} by comparing the spectrum of \yr with the Bus-DeMeo asteroid spectral catalog \citep[][]{DeMeo2009} as a test of its taxonomic spectral type. We determine the reduced $\chi^2$ statistic between the reflectance of \yr and S-complex asteroids (S, Sq, Sv, Q), C-complex (B, C, Cg, Cgh), X-complex (X, Xc, Xe, Xk, Xn), and assorted other asteroid (A, D, K, L, O, R, V). The closest match with the spectrum of \yr is to the R-type and Sa-type spectral classes with a reduced $\chi^2$ of $\sim$0.9 and 1.1. The next closest match after R-type and Sa-type asteroids is with Sr-type asteroids with a reduced $\chi^2$ of 1.7. By comparison, the reduced $\chi^2$ for the spectrum of \yr with V-types and C-types are $\sim$2.5 and 11.8.

As a rough estimate of \yrns's size, we estimate its visible albedo, $p_v$, assuming it has similar albedo properties as other small S-complex and R-type NEOs of $\sim$0.15-0.40 \citep[][]{Delbo2017}. This range contains the independent albedo estimate for \yr that we obtained by using with the NEO albedo model \citep[][]{Morbidelli2020albedo} of 0.18 when considering the $\sim$0.1 spread in albedos of small NEOs \citep[][]{Delbo2003}. We use the albedo estimate and H magnitude of \yr to determine the diameter, D, of \yr using the equation $\mathrm{D = \frac{1329}{\sqrt{p_v}}10^{-\frac{H}{5}}}$ from \citet[][]{Harris2002} arriving at D = 30-65 m. 

\begin{figure}
\centering
\includegraphics[scale=.36]{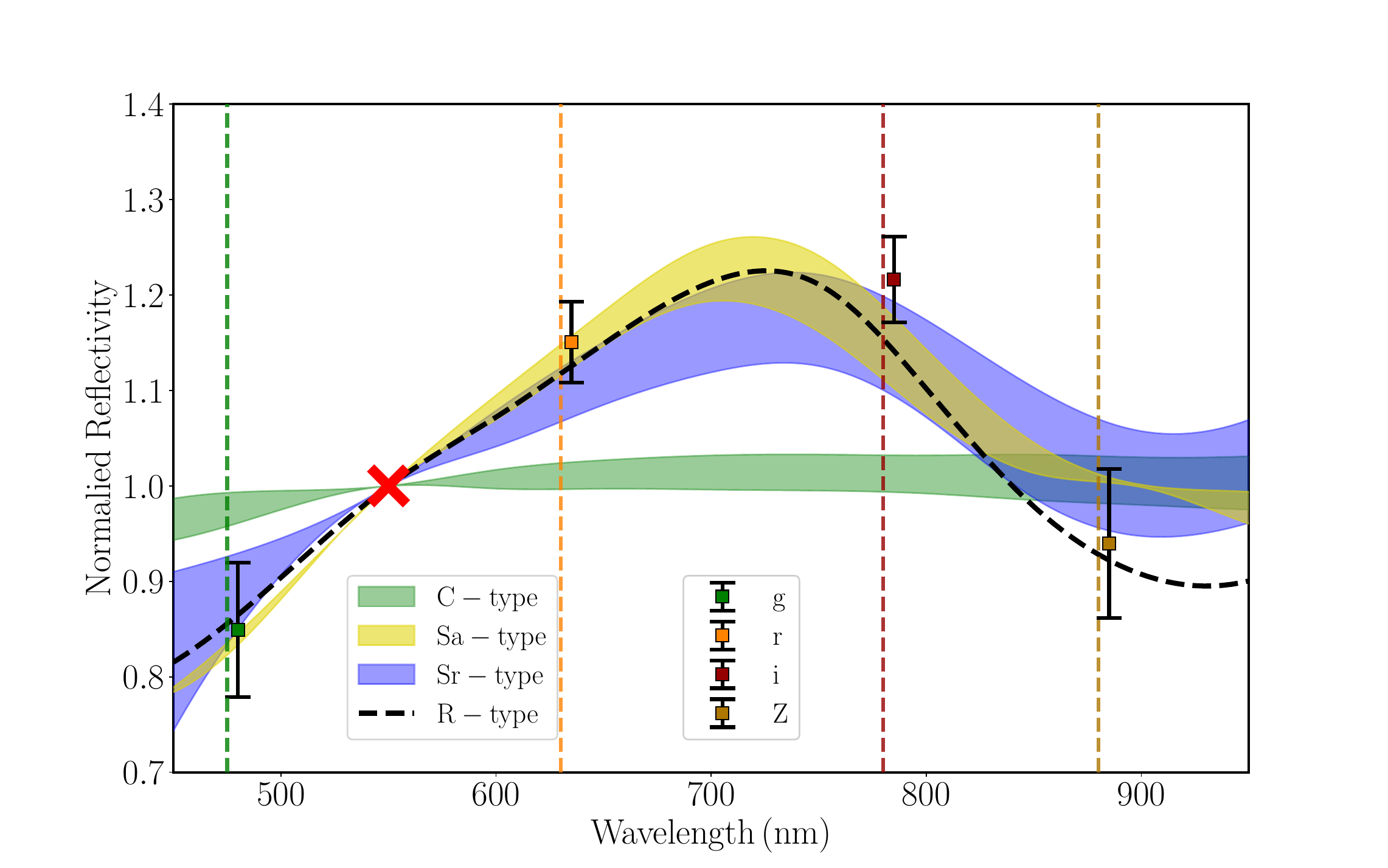}
\caption{Reflectance photometric spectrum of \yr consisting of g, r, i, and Z observations of \yr on 2024 September 27 UTC. The $\lambda_{\mathrm{eff}}$ locations of the g, r, i, and Z filters have been plotted as vertical dashed lines. The data points for the normalized reflectivity of \yr have been offset slightly from their location in the wavelength direction. The error bars on the spectrum data points correspond to 1$\sigma$ uncertainty. The red cross indicates the spectrum has been normalized to unity at 550 nm. The spectral range of R-, Sr-, Sa- and C-type asteroids from the Bus-DeMeo asteroid taxonomic catalog \citep[][]{DeMeo2009} are over-plotted with the R-type spectrum most closely resembling the spectra of \yrns. The flat C-type spectrum serves as a baseline to contrast with the other asteroid spectra.}
\end{figure}

\subsection{Lightcurve photometry and shape model}
\label{s.lightcurve}

The g, r, i, and Z GMOS photometry data of \yr from 2025 February 7 UTC were used to search for periodic variations in its brightness. The g-r, r-i, and r-Z colors computed from Section~3.2 were used to convert the g, i, and Z data into equivalent r-band magnitudes. The combined r-band equivalent g, r, i, and Z time series lightcurve are plotted in the top panel of Fig.~6. The lightcurve shows evidence of variations larger than the $\sim$0.1 mag 1-$\sigma$ photometric uncertainties of the individual data points in the lightcurve.

\begin{figure}
\centering
\includegraphics[scale=0.28]{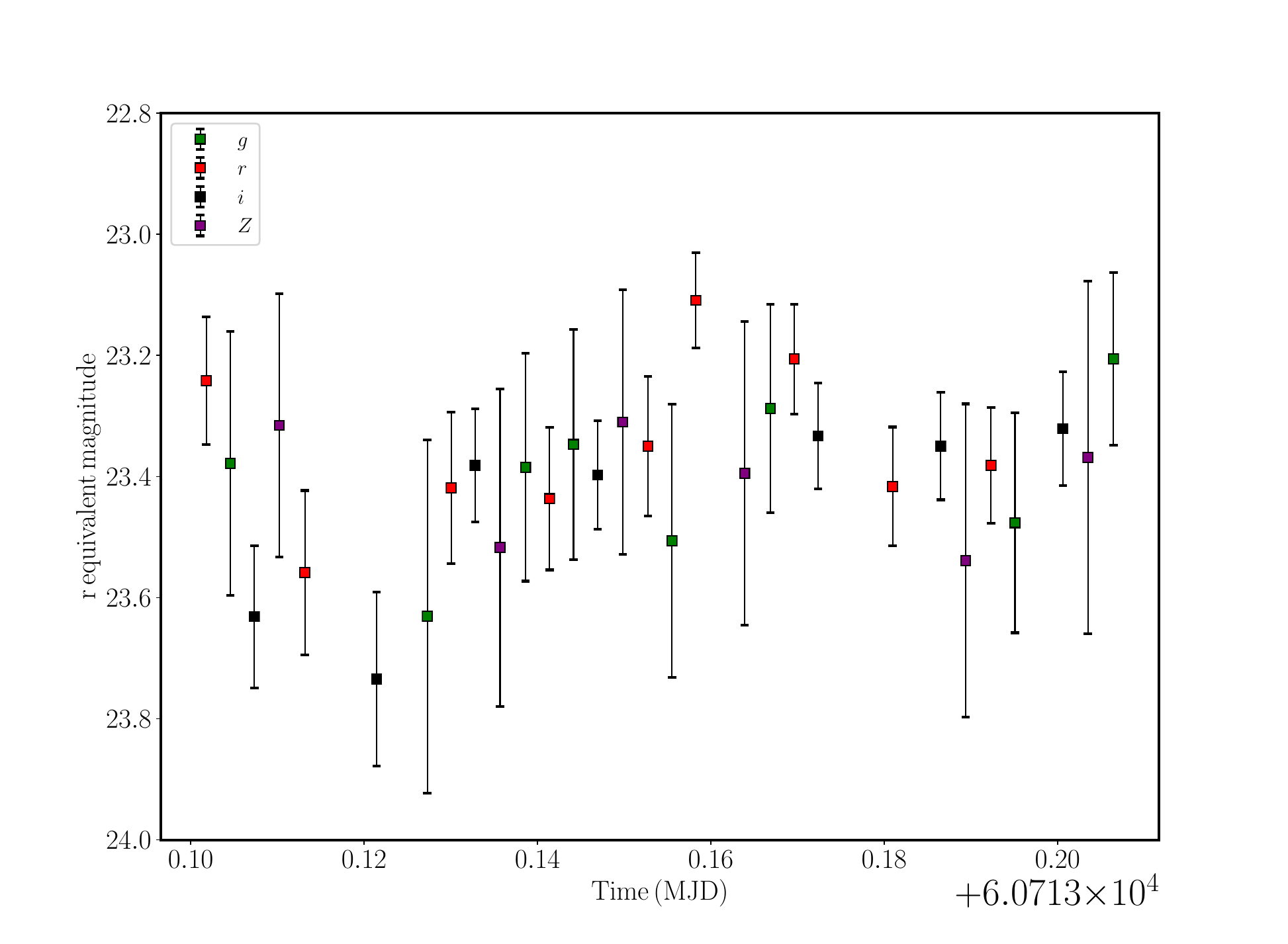}
\includegraphics[scale=0.28]{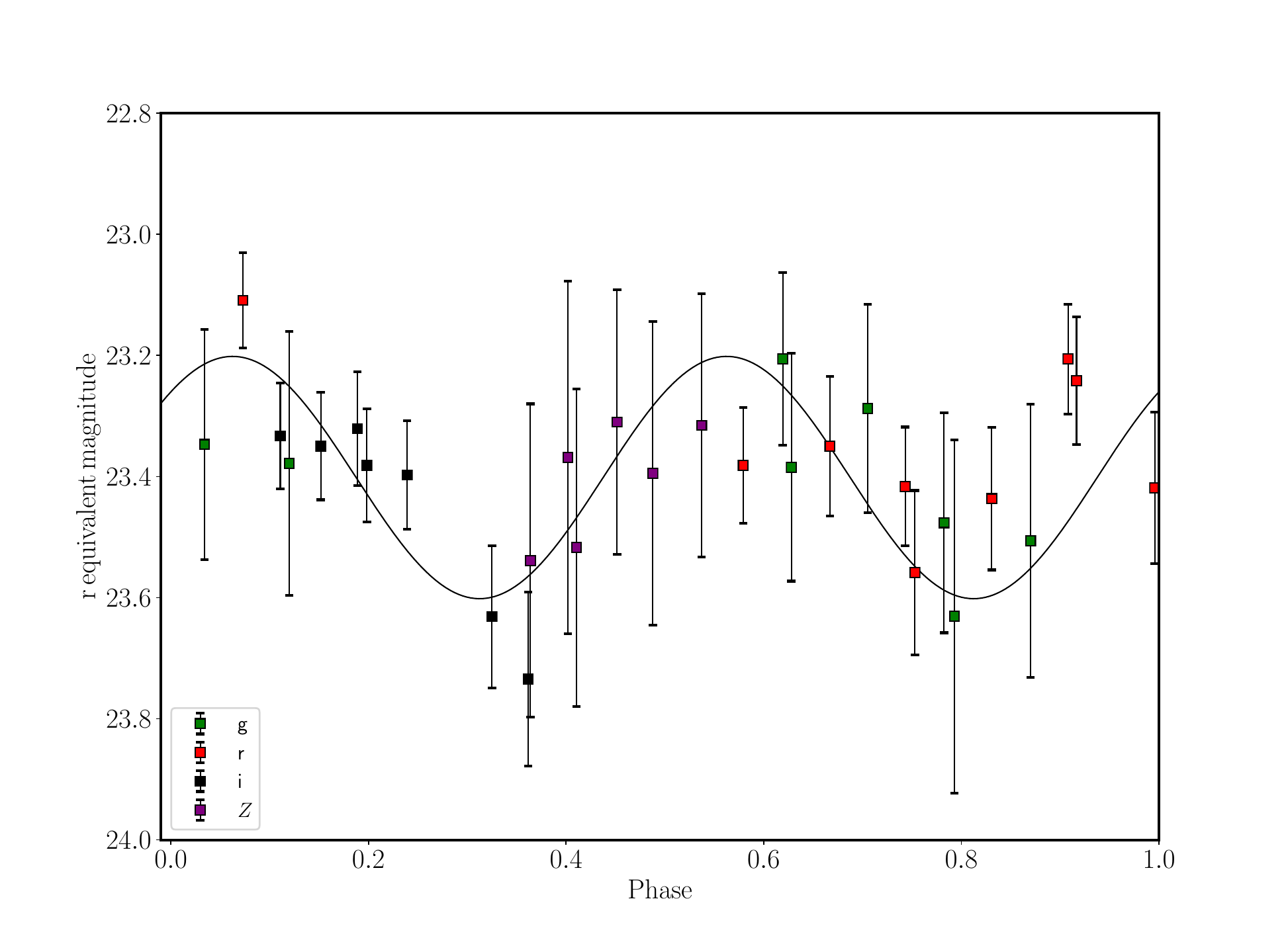}
\includegraphics[scale=0.28]{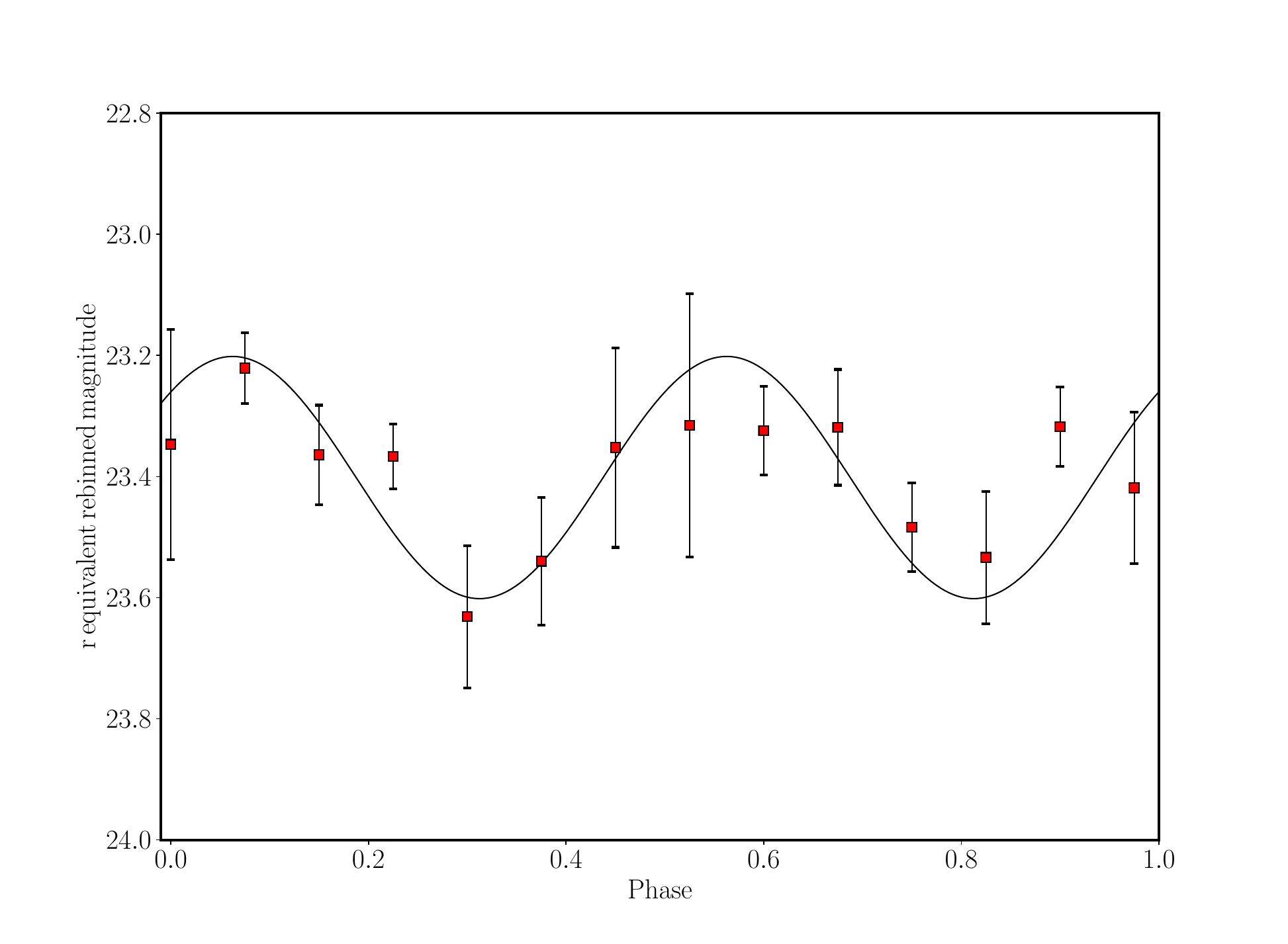}
\caption{\textbf{Top panel:} The r-band filter equivalent, time series lightcurve g, r, i, and Z observations of \yr taken by GMOS on 2025 February 7 UTC. The g, i, and Z data were converted to an equivalent r-band brightness by using the colors  g-r = 0.70$\pm$0.10, r-i = 0.25$\pm$0.06, r-Z = -0.02$\pm$0.10. The error bars on the data points equal their 1-$\sigma$ photometric uncertainties. \textbf{Center panel:} The same as in the top panel, but the data have been phased by a double-peaked rotation period of 1172 s. A sinusoudal lightcurve model with a double-peaked period of $\sim$1172 s and amplitude of 0.40 magnitudes is plotted in black. \textbf{Bottom panel:} the same as in the center panel, except that the data have been re-binned in with 0.075 wide bins in phase dimensional space.}
\end{figure}

We take a similar approach as described by \citet[][]{Bolin2024Streak} to estimating the lightcurve period of the r-band equivalent data by applying a Lomb-Scargle periodogram \citep[LS,][]{Lomb1976,Scargle1982} algorithm and a phase dispersion minimization algorithm \citep[PDM, ][]{Stellingwerf1978} to the data. The LS and PDM algorithms result in finding an estimated lightcurve period of $\sim$587-590s as seen in the top and bottom panels of Fig.~7. We use the approach of \citep[][]{Purdum2021} estimate the uncertainty on the determined period by removing $\sqrt{N}$ data points from the time series lightcurve and repeating our periodogram estimation of the lightcurve period $\sim$1,000 times, resulting in a central value of $\sim$586 s and a 1~$\sigma$ uncertainty estimate of $\sim$142 s. These period estimates imply that \yr has a double-peaked synodic rotation period of $\sim$1172$\pm$284~s.

The central panel of Fig.~6 shows the r-band equivalent lightcurve of \yr folded by the double-peaked rotation period of 1172 s. The result shows almost complete coverage in the lightcurve of \yr over its 1172 s rotation. We rebin the phased lightcurve of \yr into 0.075 wide bins in phase dimensional space as seen in the bottom panel of Fig.~6. We note that estimating an asteroid's shape from a single lightcurve observation is affected by the degeneracy between viewing angle and axial ratio \citep[e.g.,][]{Bolin2018}, therefore, we will take a multi-epoch approach to estimating the asteroid's shape as described below.

\begin{figure}
\centering
\includegraphics[scale=0.28]{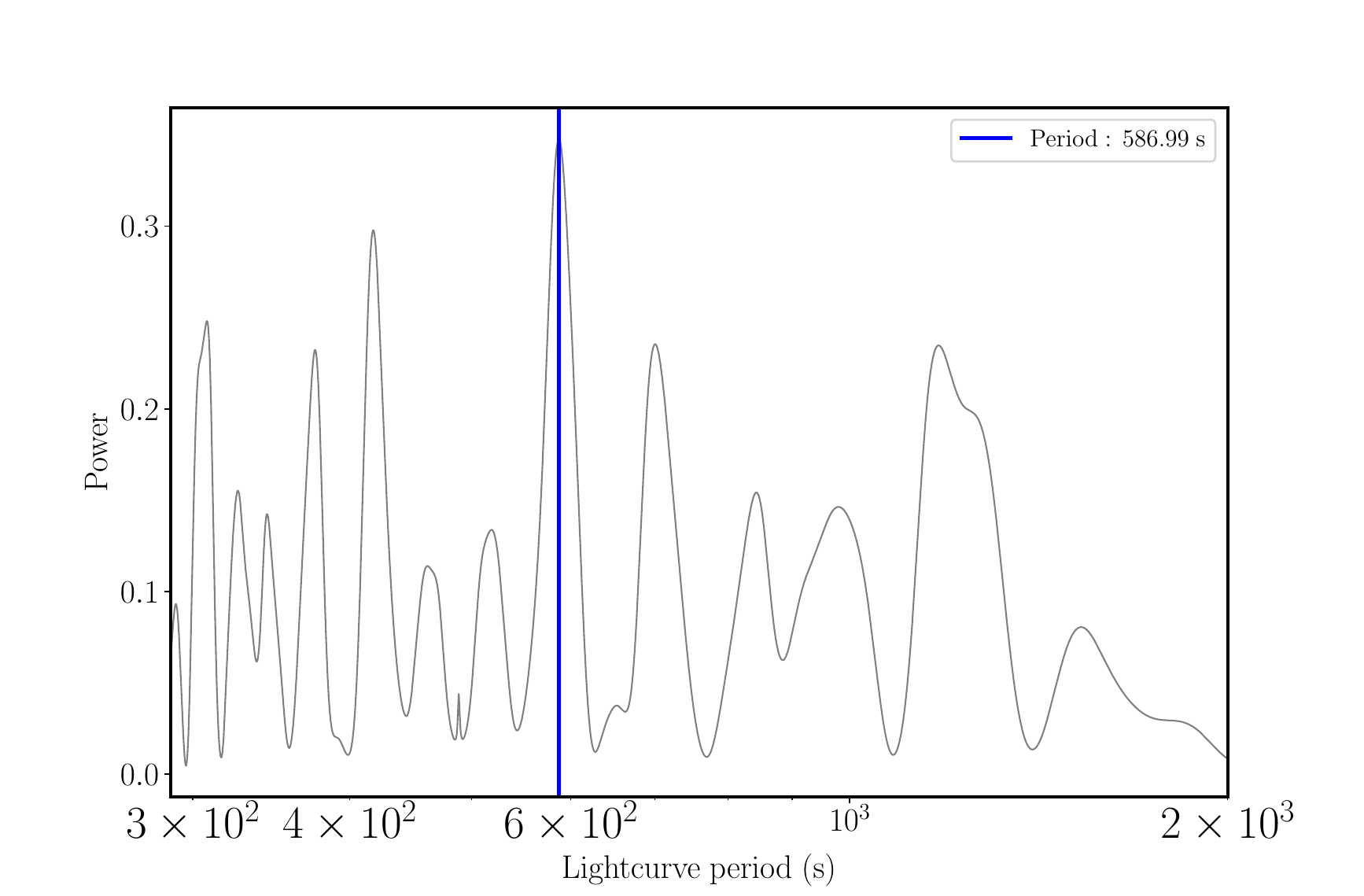}
\includegraphics[scale=0.28]{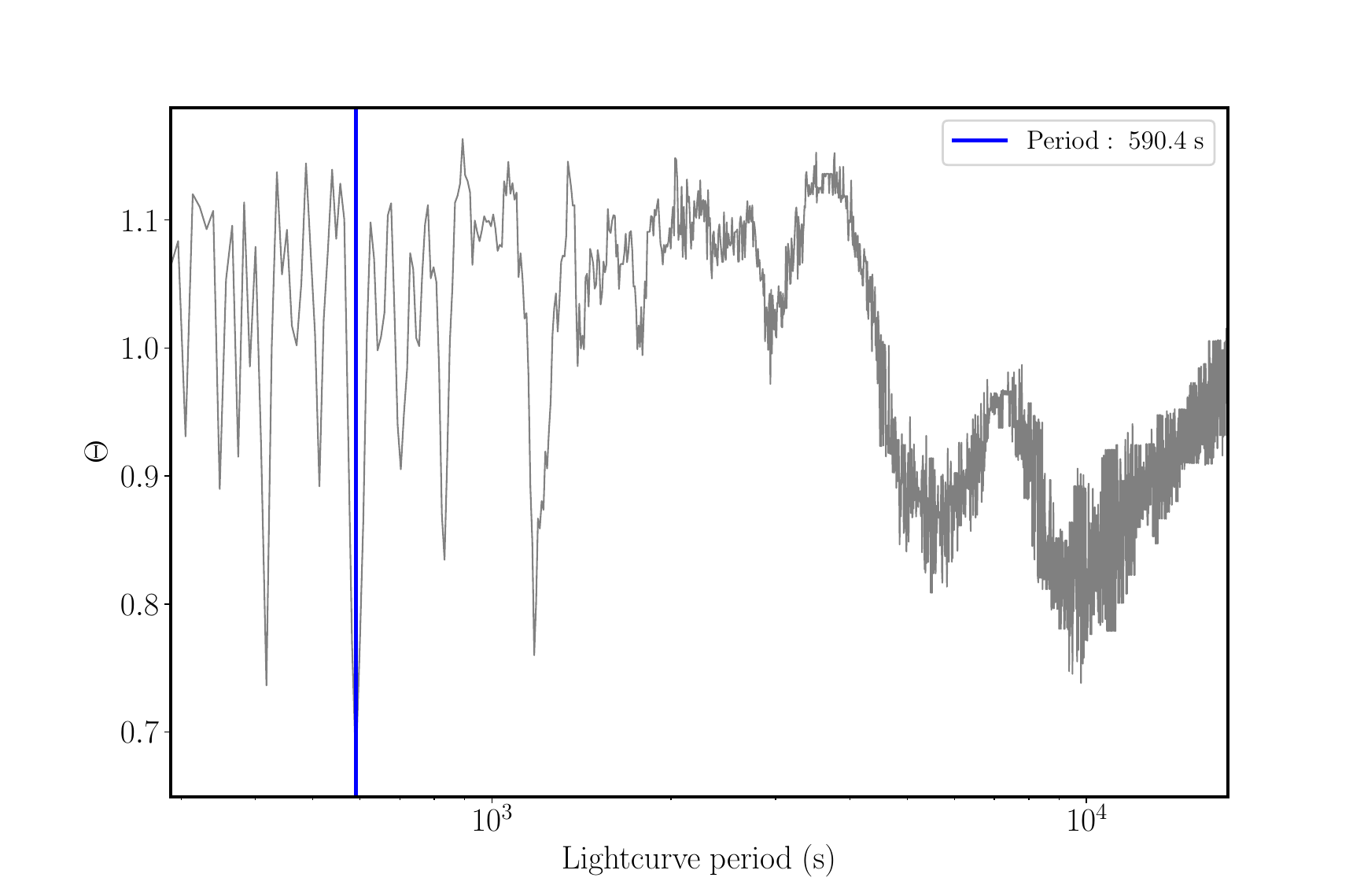}
\caption{\\textbf{Left panel:} Lomb-Scargle periodogram of lightcurve period vs. spectral power \citep[][]{Lomb1976} for the g filter data from the 2024 September 27 UTC Gemini N/GMOS observations. A peak in power is located at a single-peaked lightcurve period of 587 s. \textbf{Right panel:} Phase dispersion minimization analysis of lightcurve rotation period vs. $\Theta$ metric \citep[][]{Stellingwerf1978}. The $\Theta$ metric is minimized at single-peaked rotation periods of 1150 s, similar to the 1320$\pm$227 s period found with the Lomb-Scargle Periodogram.}
\label{fFiig:periodogram}
\end{figure}

To estimate a more accurate shape for \yrns, we use the lightcurve inversion method \citep{Kaasalainen2001b, Kaasalainen2001a}, which has been widely applied to derive the physical properties of small bodies, including near-Earth asteroids (NEAs). A similar approach was successfully used in the case of (3200) Phaethon, where photometric data from multiple observing geometries determined its rotational state and shape \citep{Hanus20183200}.

For \yrns, we used a combination of dense lightcurves obtained from Gemini (MPC observatory code, I11) we used only data in the r and g filters that are the most accurate), public data taken by the Very Large Telescope (MPC observatory code 309), and sparse photometry reported to the Minor Planet Center (MPC) that provided sufficient coverage of observing geometries. The modeling procedure followed the methodology outlined in our previous studies \citep{Athanasopoulos2022,Hanus2023}.

We used photometric observations reported to the MPC selecting data from observatories providing magnitude measurements with at least two decimal places were considered to ensure the highest possible accuracy. After initial modeling attempts, we further filtered out the MPC photometry, ending with three data sets. Namely, observations from multiple ATLAS \citep[Asteroid Terrestrial-impact Last Alert System;][]{Tonry2018} stations (MPC observatory codes (T05, T08, and W68) were combined into a single data set in the o filter, consisting of 25 individual measurements. Additional observations from the Catalina Sky Survey \citep{Larson2003} (MPC observatory code 703), Pan-STARRS2 (MPC observatory code F52) \citep{Chambers2016} stations combined together in the G filter finalized the sparse dataset.

The photometric data were processed following the methodology outlined in \citet{Hanus2011} to prepare them for lightcurve inversion. This procedure included transformation to fluxes, epoch light-travel time correction, flux normalization to unit distance to the Sun and the observer, and unity. Any outliers or measurements affected by systematic errors were identified and removed.

The additional dense photometric observations of \yr were obtained from the European Southern Observatory (ESO) Science Archive as part of a long-term monitoring program (ESO Program ID 113.2690.002, PI Hainaut). The program, conducted in Service Mode at VLT UT1-Antu, supports the ESA Space Situational Awareness Near-Earth Objects Protection Program, which aims to characterize the physical properties of potentially hazardous asteroids. The observations were taken with the FORS2 instrument on 2025 January 21 UTC.

The VLT data are publicly available, and we retrieved them from the ESO database as raw images. The data were processed using the MPO Canopus software package, following standard photometric reduction procedures \citep[][]{Warner2009}. Bias and flat-field corrections were applied to all frames, followed by aperture photometry extraction, calibrated against known field stars to ensure accurate magnitude measurements.

The available photometric dataset spans 42 days of observations, covering a range of phase angles from 45.7$^\circ$ to 14.8$^\circ$, which is critical for constraining the shape model. Table A.1 summarizes the complete list of photometric measurements, including observation epochs, filter bands, and sources.

\begin{figure}
    \centering
    \includegraphics[width=0.7\linewidth]{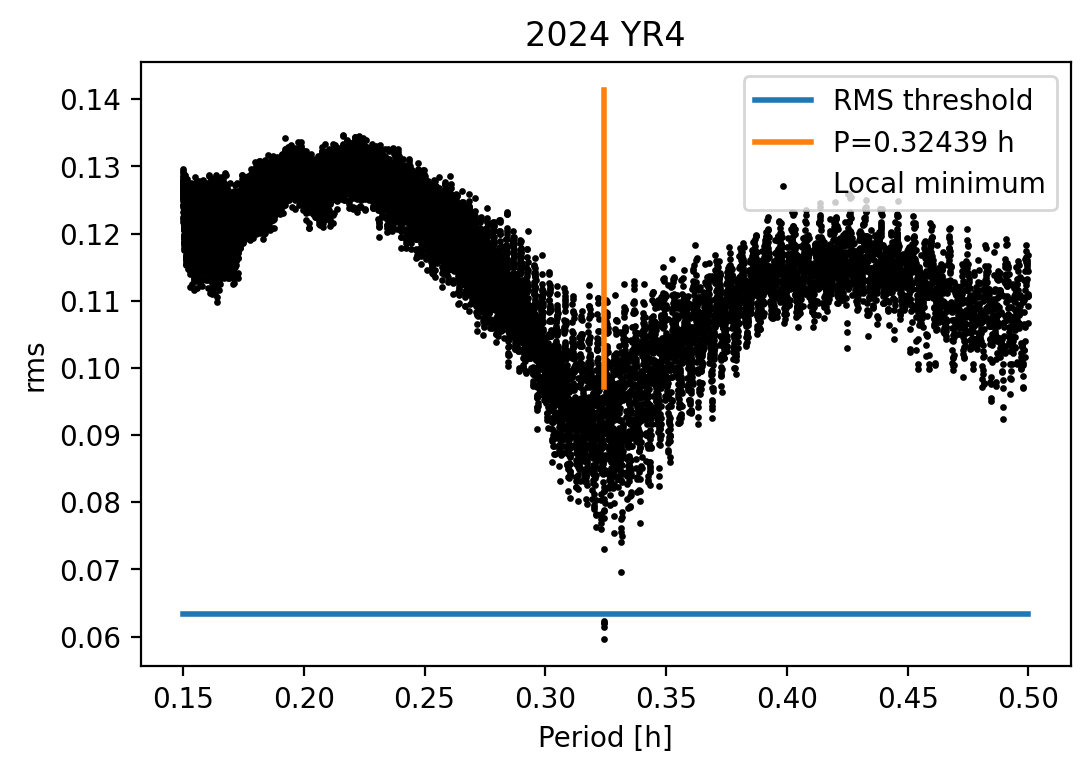}
    \caption{The periodogram of 2024 YR4, computed near the suspected synodic period. The minimum at $P_{\mathrm{sid}} = 0.32439$ hours, indicated by the vertical line, corresponds to the global minimum. The horizontal line represents the RMS threshold, indicating the statistical significance of the minimum.}
    \label{fig:periodogram}
\end{figure}

We conducted a broad period search within the range of 0.15 to 0.50 hours, encompassing the suspected synodic period of approximately $\sim$0.32 hours derived from the photometric dataset. A well-defined minimum was identified in the periodogram at $P_{\mathrm{sid}} = 0.32439$ hours (1168s) (Figure~\ref{fig:periodogram}). This solution remains stable against variations in the relative weighting of the individual datasets, confirming its robustness.

Adopting this sidereal period, we explored possible spin-vector orientations by testing a wide range of initial values isotropically distributed on a sphere. The results indicate a degree of sensitivity to the chosen data weighting. Several pole solutions were tested using different relative weight configurations, with the most prominent solution found at an ecliptic longitude and latitude of approximately $\lambda = 42^\circ$ and $\beta = -25^\circ$, respectively. This particular solution satisfies physical constraints, ensuring that the derived shape model rotates around its principal axis with maximum momentum of inertia. While alternative pole solutions cannot be entirely excluded, this solution appears to be the most plausible. This solution will be included in the Database of Asteroid Models from Inversion Techniques (DAMIT) lightcurve inversion software package \citep[][]{Durech2015}.

\begin{figure}
    \centering
    \resizebox{0.99\hsize}{!}{\includegraphics{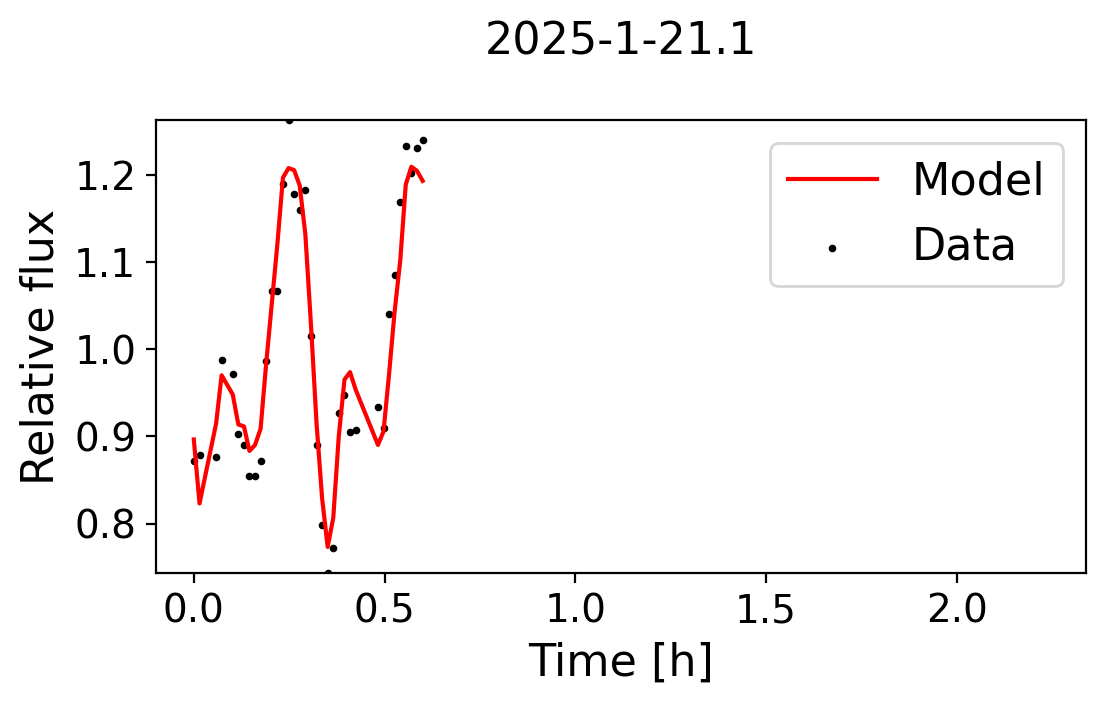}\includegraphics{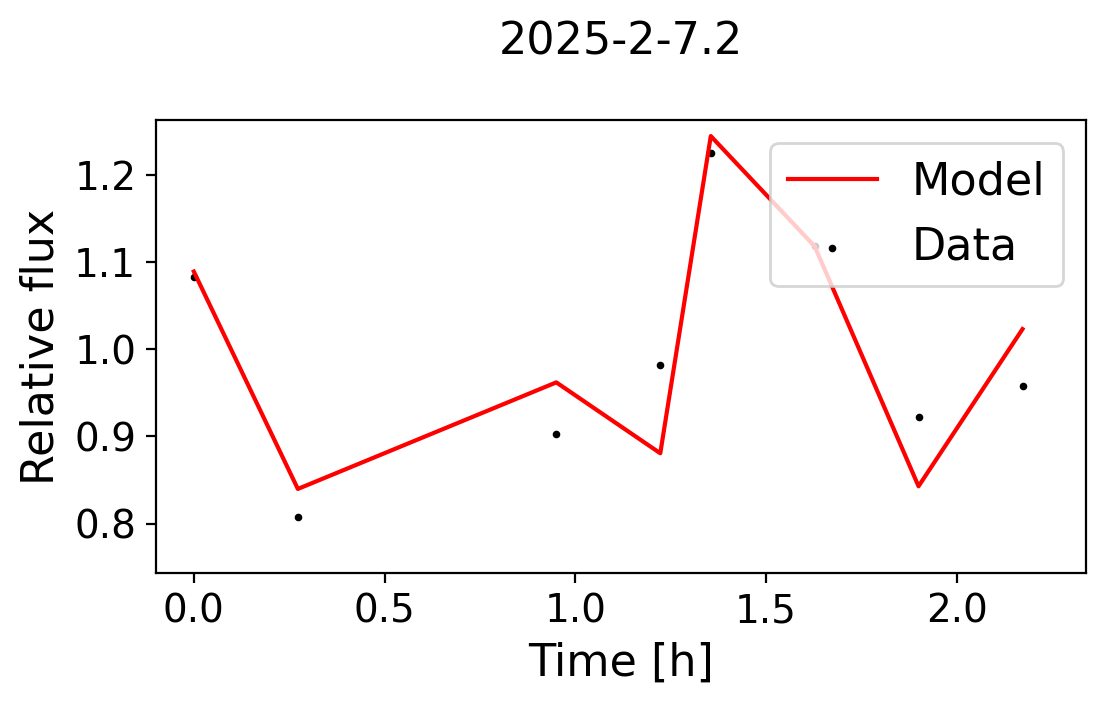}\includegraphics{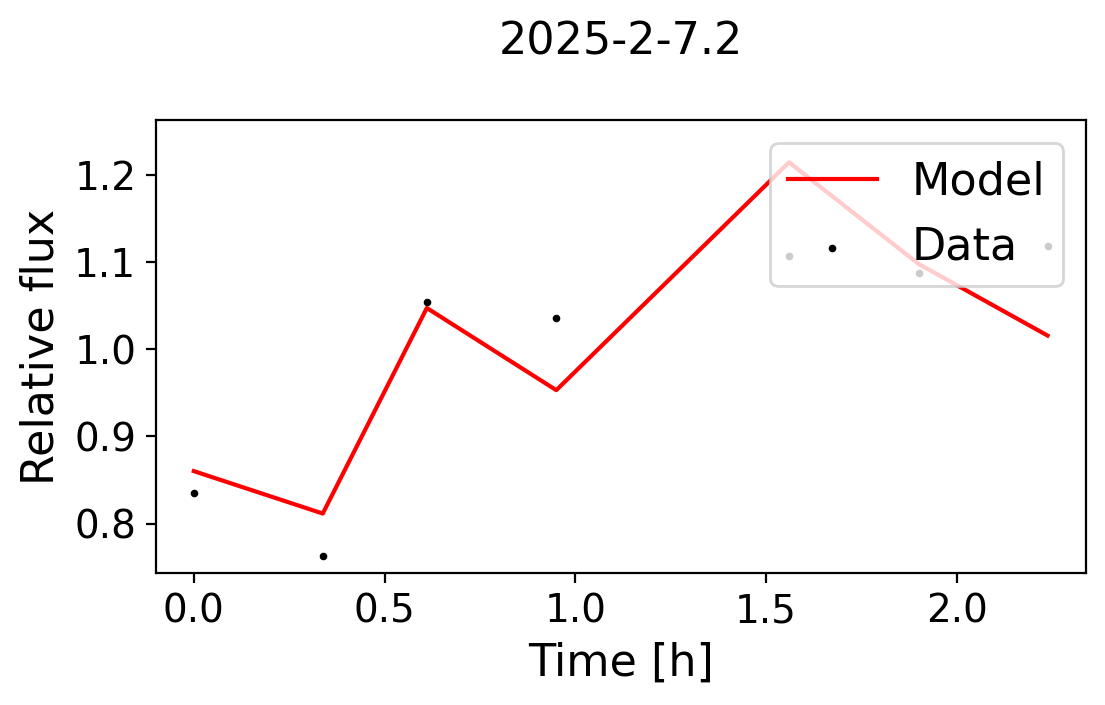}}
    \includegraphics[width=0.99\linewidth]{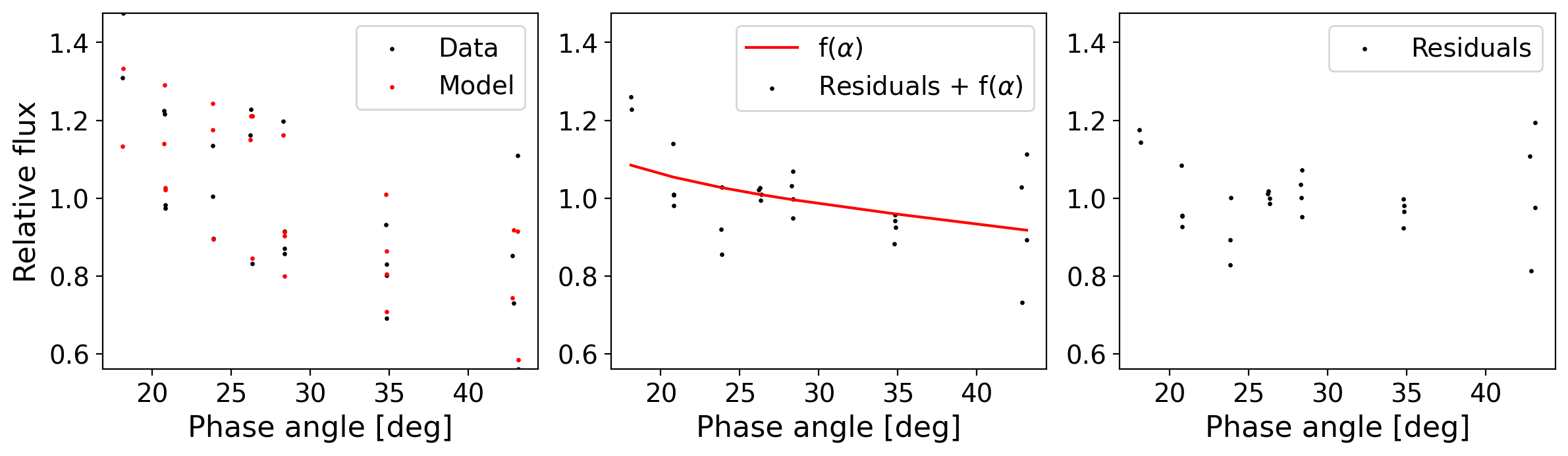}
    \includegraphics[width=0.99\linewidth]{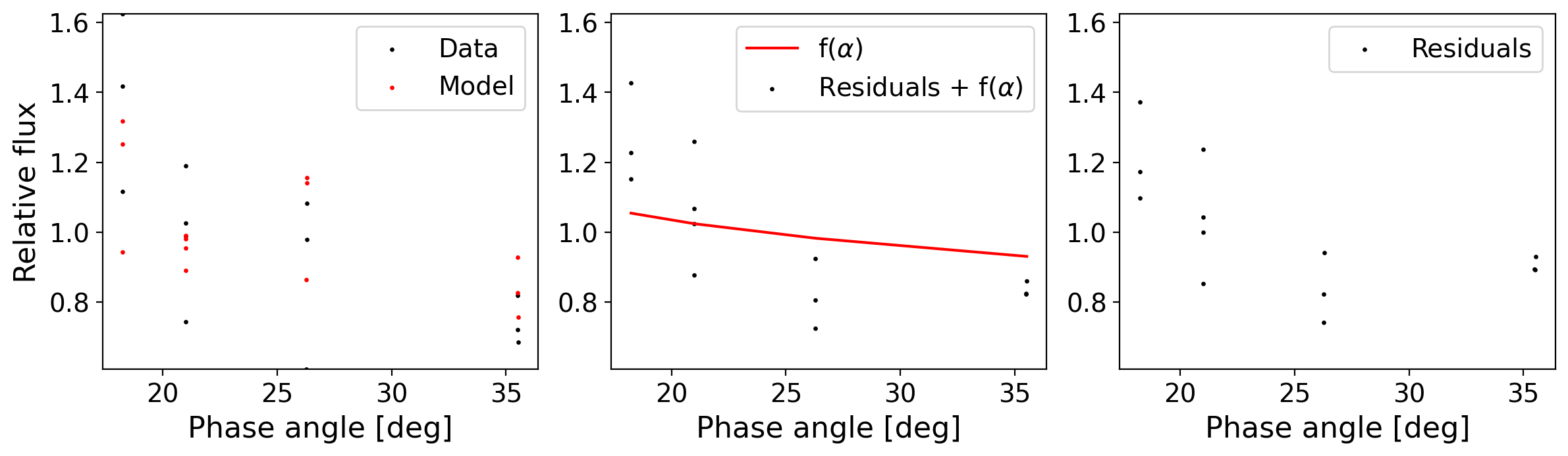}
    \includegraphics[width=0.99\linewidth]{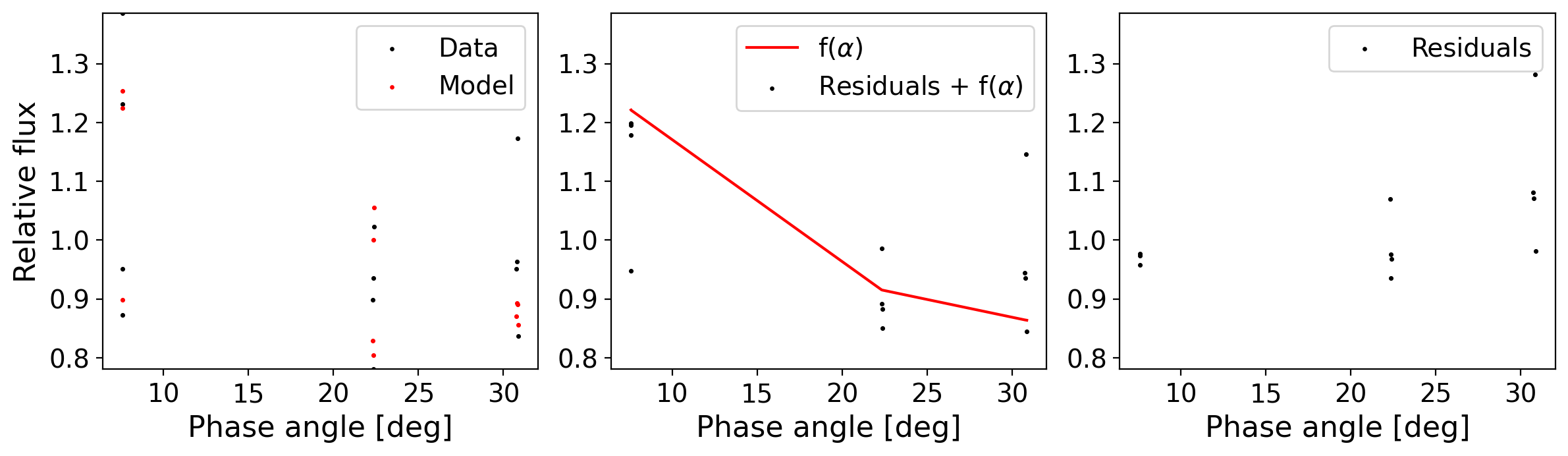}
    \caption{Lightcurve fit of 2024 YR4, showing dense photometric observations from VLT and Gemini in the \textit{r} and \textit{g} filters (top panels), and sparse datasets from ATLAS, Catalina Sky Survey, and Pan-STARRS (bottom three panels). The best-fit model is overlaid.}
    \label{fig:lcs_fit}
\end{figure}

Figure~\ref{fig:lcs_fit} presents the fit to the photometric data, incorporating dense lightcurves obtained from VLT and Gemini in the \textit{r} and \textit{g} filters, along with sparse photometric measurements from ATLAS, the Catalina Sky Survey, and Pan-STARRS. The derived shape of 2024 YR4 is notably peculiar, resembling a highly flattened spheroid with nearly equal equatorial dimensions (Fig.~\ref{fig:periodogram}).

\begin{figure}
\centering
\includegraphics[scale=0.38]{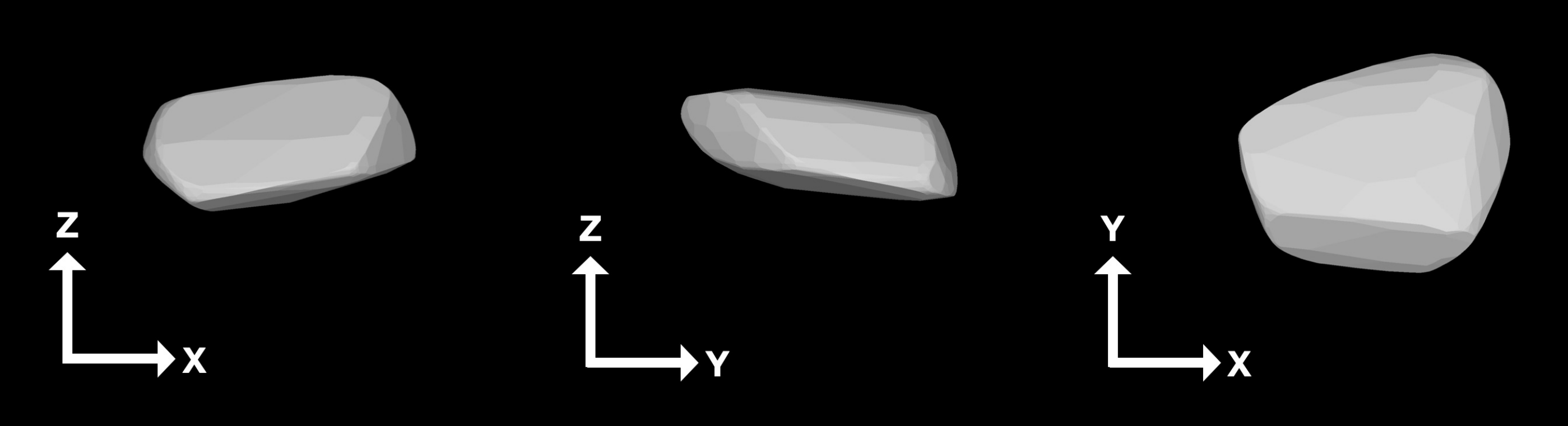}
\caption{Convex shape model of \yr in three orthogonal views obtained using publicly available photometry of \yr from the Minor Planet Center archive, VLT archive and Gemini. The X vs. Z and Y vs. Z views show the edge on view of the asteroid with the spin axis pointing in the Z direction (ecliptic longitude and latitude equal to 42$^{\circ}$ and -25 degrees$^{\circ}$). The asteroid is viewed from a pole-on view in the X vs. Y view.}
\label{fFiig:periodogram}
\end{figure}

\section{Discussion and Conclusions}

The properties of \yr are similar to other small NEOs, has a R-type/S-complex composition, most likely originated from a resonances located inner Main Belt, and likely has moderate reflective properties implying a size of $\sim$42 m. The exact origin within the Main Belt seems less clear, though a clue may lie in its spin-vector orientation. As recent asteroid lightcurve spin-vector studies have shown, the spin direction of an asteroid can affect the way an asteroid's orbit evolves due to the thermal recoil Yarkovsky effect, with prograde asteroids drifting outward, and retrograde asteroids drifting inwards \citep[][]{Bolin2018size,Hanus2018SpinVectorEos,Athanasopoulos2022}. Since the spin direction of \yr is retrograde with its -39$^{\circ}$ ecliptic latitude, an origin from the inner Main Belt since its retrograde spin would cause it to drift inwards away from the 3:1 MMR at 2.5 au, its most likely escape path into the NEO population. 

Therefore, \yrns's suspected retrograde spin may imply that it original location was the central Main Belt, located at 2.52 au to 2.82 au from the Sun \citep[][]{Nesvorny2015a}. This would take \yr's origins away from the inner Main Belt where S-complex asteroids dominate to the central Main Belt where C-complex asteroids are more plentiful seemingly in tension with its taxonomic type \citep[][]{DeMeo2014a}. However, we caution that several S-complex families are located near the 3:1 MMR in the central Main Belt, such as the Rafita family located at 2.58 au, which could be a candidate source family for \yrns. In addition, recent lightcurve studies of asteroid families show that some asteroid can be located on the exterior side of an asteroid family's V-shape and still have a retrograde spin \citep[][]{Athanasopoulos2024}. We expect more ECAs to be found in the future by surveys targeting NEAs \citep[e.g.,][]{Larson1998,Tonry2018}, or in large, all-sky surveys \citep[e.g.,][]{Whidden2019,Schwamb2023ApJS,Masiero2024}.

\section*{acknowledgments}

This study is based on observations obtained at the international Gemini Observatory, a program of NSF's NOIRLab, which is managed by the Association of Universities for Research in Astronomy (AURA) under a cooperative agreement with the National Science Foundation on behalf of the Gemini Observatory partnership.

Keck Observatory is located on Maunakea, land of the K$\mathrm{\bar{a}}$naka Maoli people, and a mountain of considerable cultural, natural, and ecological significance to the indigenous Hawaiian people. The authors wish to acknowledge the importance and reverence of Maunakea and express gratitude for the opportunity to conduct observations from the mountain.

This work has been supported by the grant 25-16789S of the Czech Science Foundation.

The authors wish to thank Matthew Belyakov for help with the observaitons of \yrns, and Mauro Barbieri for help with searching the VLT data archive. The authors also wish to thank K. Wierzchos, and C. Fuls for their help with measuring the astrometry from the Gemini images and submitting it to the MPC. The authors also wish to think the team members of the Catalina Sky Survey, the Panoramic Survey Telescope and Rapid Response System, an ATLAS for submitting observations of \yr to the MPC which were used in this work.

This research has made use of data and/or services provided by the International Astronomical Union, Minor Planet Center.

\facility{ATLAS, Gemini:South, Keck:I, VLT:Antu} 

\bibliographystyle{aasjournal}
\bibliography{/Users/bolin/Dropbox/Projects/latex_references_commands_bib_styles/neobib.bib}

\begin{thebibliography}{}
\expandafter\ifx\csname natexlab\endcsname\relax\def\natexlab#1{#1}\fi
\providecommand{\url}[1]{\href{#1}{#1}}
\providecommand{\dodoi}[1]{doi:~\href{http://doi.org/#1}{\nolinkurl{#1}}}
\providecommand{\doeprint}[1]{\href{http://ascl.net/#1}{\nolinkurl{http://ascl.net/#1}}}
\providecommand{\doarXiv}[1]{\href{https://arxiv.org/abs/#1}{\nolinkurl{https://arxiv.org/abs/#1}}}

\bibitem[{{Athanasopoulos} {et~al.}(2022){Athanasopoulos}, {Hanu{\v{s}}},
  {Avdellidou}, {Bonamico}, {Delbo}, {Conjat}, {Ferrero}, {Gazeas}, {Rivet},
  {Sioulas}, {van Belle}, {Antonini}, {Audejean}, {Behrend}, {Bernasconi},
  {Brinsfield}, {Brouillard}, {Brunetto}, {Fauvaud}, {Fauvaud}, {Gonzalez},
  {Higgins}, {Holoien}, {Kober}, {Koff}, {Kryszczynska}, {Livet}, {Marciniak},
  {Oey}, {Pejcha}, {Rives}, \& {Roy}}]{Athanasopoulos2022}
{Athanasopoulos}, D., {Hanu{\v{s}}}, J., {Avdellidou}, C., {et~al.} 2022, \aap,
  666, A116, \dodoi{10.1051/0004-6361/202243905}

\bibitem[{{Athanasopoulos} {et~al.}(2024){Athanasopoulos}, {Hanu{\v{s}}},
  {Avdellidou}, {van Belle}, {Ferrero}, {Bonamico}, {Gazeas}, {Delbo}, {Rivet},
  {Apostolovska}, {Todorovi{\'c}}, {Novakovic}, {Bebekovska}, {Romanyuk},
  {Bolin}, {Zhou}, \& {Agrusa}}]{Athanasopoulos2024}
---. 2024, \aap, 690, A215, \dodoi{10.1051/0004-6361/202451363}

\bibitem[{{Avdellidou} {et~al.}(2022){Avdellidou}, {Delbo}, {Morbidelli},
  {Walsh}, {Munaibari}, {Bourdelle de Micas}, {Devog{\`e}le}, {Fornasier},
  {Gounelle}, \& {van Belle}}]{Avdellidou2022}
{Avdellidou}, C., {Delbo}, M., {Morbidelli}, A., {et~al.} 2022, \aap, 665, L9,
  \dodoi{10.1051/0004-6361/202244590}

\bibitem[{{Binzel} {et~al.}(2015){Binzel}, {Reddy}, \& {Dunn}}]{Binzel2015}
{Binzel}, R.~P., {Reddy}, V., \& {Dunn}, T.~L. 2015, {The Near-Earth Object
  Population: Connections to Comets, Main-Belt Asteroids, and Meteorites},
  243--256

\bibitem[{{Binzel} {et~al.}(2010){Binzel}, {Morbidelli}, {Merouane}, {DeMeo},
  {Birlan}, {Vernazza}, {Thomas}, {Rivkin}, {Bus}, \& {Tokunaga}}]{Binzel2010}
{Binzel}, R.~P., {Morbidelli}, A., {Merouane}, S., {et~al.} 2010, \nat, 463,
  331, \dodoi{10.1038/nature08709}

\bibitem[{{Binzel} {et~al.}(2019){Binzel}, {DeMeo}, {Turtelboom}, {Bus},
  {Tokunaga}, {Burbine}, {Lantz}, {Polishook}, {Carry}, {Morbidelli}, {Birlan},
  {Vernazza}, {Burt}, {Moskovitz}, {Slivan}, {Thomas}, {Rivkin}, {Hicks},
  {Dunn}, {Reddy}, {Sanchez}, {Granvik}, \& {Kohout}}]{Binzel2019}
{Binzel}, R.~P., {DeMeo}, F.~E., {Turtelboom}, E.~V., {et~al.} 2019, \icarus,
  324, 41, \dodoi{10.1016/j.icarus.2018.12.035}

\bibitem[{{Bolin} {et~al.}(2025){Bolin}, {Denneau}, {Abron}, {Jedicke},
  {Chiboucas}, {Ingebretsen}, \& {Lemaux}}]{Bolin2025PT5}
{Bolin}, B.~T., {Denneau}, L., {Abron}, L.-M., {et~al.} 2025, \apjl, 978, L37,
  \dodoi{10.3847/2041-8213/ada1d0}

\bibitem[{{Bolin} {et~al.}(2024){Bolin}, {Ghosal}, \&
  {Jedicke}}]{Bolin2024Streak}
{Bolin}, B.~T., {Ghosal}, M., \& {Jedicke}, R. 2024, \mnras, 527, 1633,
  \dodoi{10.1093/mnras/stad3227}

\bibitem[{{Bolin} {et~al.}(2018{\natexlab{a}}){Bolin}, {Morbidelli}, \&
  {Walsh}}]{Bolin2018size}
{Bolin}, B.~T., {Morbidelli}, A., \& {Walsh}, K.~J. 2018{\natexlab{a}}, \aap,
  611, A82, \dodoi{10.1051/0004-6361/201732079}

\bibitem[{{Bolin} {et~al.}(2023){Bolin}, {Noll}, {Caiazzo}, {Fremling}, \&
  {Binzel}}]{Bolin2023Dink}
{Bolin}, B.~T., {Noll}, K.~S., {Caiazzo}, I., {Fremling}, C., \& {Binzel},
  R.~P. 2023, \icarus, 400, 115562, \dodoi{10.1016/j.icarus.2023.115562}

\bibitem[{{Bolin} {et~al.}(2018{\natexlab{b}}){Bolin}, {Weaver}, {Fernandez},
  {Lisse}, {Huppenkothen}, {Jones}, {Juri{\'c}}, {Moeyens}, {Schambeau},
  {Slater}, {Ivezi{\'c}}, \& {Connolly}}]{Bolin2018}
{Bolin}, B.~T., {Weaver}, H.~A., {Fernandez}, Y.~R., {et~al.}
  2018{\natexlab{b}}, \apjl, 852, L2, \dodoi{10.3847/2041-8213/aaa0c9}

\bibitem[{{Bolin} {et~al.}(2020){Bolin}, {Fremling}, {Holt}, {Hankins},
  {Ahumada}, {Anand}, {Bhalerao}, {Burdge}, {Copperwheat}, {Coughlin},
  {Deshmukh}, {De}, {Kasliwal}, {Morbidelli}, {Purdum}, {Quimby}, {Bodewits},
  {Chang}, {Ip}, {Hsu}, {Laher}, {Lin}, {Lisse}, {Masci}, {Ngeow}, {Tan},
  {Zhai}, {Burruss}, {Dekany}, {Delacroix}, {Duev}, {Graham}, {Hale},
  {Kulkarni}, {Kupfer}, {Mahabal}, {Mr{\'o}z}, {Neill}, {Riddle}, {Rodriguez},
  {Smith}, {Soumagnac}, {Walters}, {Yan}, \& {Zolkower}}]{Bolin2020CD3}
{Bolin}, B.~T., {Fremling}, C., {Holt}, T.~R., {et~al.} 2020, \apjl, 900, L45,
  \dodoi{10.3847/2041-8213/abae69}

\bibitem[{{Bolin} {et~al.}(2021){Bolin}, {Fernandez}, {Lisse}, {Holt}, {Lin},
  {Purdum}, {Deshmukh}, {Bauer}, {Bellm}, {Bodewits}, {Burdge}, {Carey},
  {Copperwheat}, {Helou}, {Ho}, {Horner}, {van Roestel}, {Bhalerao}, {Chang},
  {Chen}, {Hsu}, {Ip}, {Kasliwal}, {Masci}, {Ngeow}, {Quimby}, {Burruss},
  {Coughlin}, {Dekany}, {Delacroix}, {Drake}, {Duev}, {Graham}, {Hale},
  {Kupfer}, {Laher}, {Mahabal}, {Mr{\'o}z}, {Neill}, {Riddle}, {Rodriguez},
  {Smith}, {Soumagnac}, {Walters}, {Yan}, \& {Zolkower}}]{Bolin2021LD2}
{Bolin}, B.~T., {Fernandez}, Y.~R., {Lisse}, C.~M., {et~al.} 2021, \aj, 161,
  116, \dodoi{10.3847/1538-3881/abd94b}

\bibitem[{{Bolin} {et~al.}(2022){Bolin}, {Ahumada}, {van Dokkum}, {Fremling},
  {Granvik}, {Hardegree-Ullman}, {Harikane}, {Purdum}, {Serabyn}, {Southworth},
  \& {Zhai}}]{Bolin2022IVO}
{Bolin}, B.~T., {Ahumada}, T., {van Dokkum}, P., {et~al.} 2022, \mnras, 517,
  L49, \dodoi{10.1093/mnrasl/slac089}

\bibitem[{{Bowell} {et~al.}(1988){Bowell}, {Hapke}, {Domingue}, {Lumme},
  {Peltoniemi}, \& {Harris}}]{Bowell1988}
{Bowell}, E., {Hapke}, B., {Domingue}, D., {et~al.} 1988, Asteroids II, 399

\bibitem[{{Bus} \& {Binzel}(2002)}]{Bus2002}
{Bus}, S.~J., \& {Binzel}, R.~P. 2002, Icarus, 158, 146,
  \dodoi{10.1006/icar.2002.6856}

\bibitem[{{Casali} {et~al.}(2007){Casali}, {Adamson}, {Alves de Oliveira},
  {Almaini}, {Burch}, {Chuter}, {Elliot}, {Folger}, {Foucaud}, {Hambly},
  {Hastie}, {Henry}, {Hirst}, {Irwin}, {Ives}, {Lawrence}, {Laidlaw}, {Lee},
  {Lewis}, {Lunney}, {McLay}, {Montgomery}, {Pickup}, {Read}, {Rees}, {Robson},
  {Sekiguchi}, {Vick}, {Warren}, \& {Woodward}}]{Casali2007}
{Casali}, M., {Adamson}, A., {Alves de Oliveira}, C., {et~al.} 2007, \aap, 467,
  777, \dodoi{10.1051/0004-6361:20066514}

\bibitem[{{Chambers} {et~al.}(2016){Chambers}, {Magnier}, {Metcalfe},
  {Flewelling}, {Huber}, {Waters}, {Denneau}, {Draper}, {Farrow}, {Finkbeiner},
  {Holmberg}, {Koppenhoefer}, {Price}, {Saglia}, {Schlafly}, {Smartt},
  {Sweeney}, {Wainscoat}, {Burgett}, {Grav}, {Heasley}, {Hodapp}, {Jedicke},
  {Kaiser}, {Kudritzki}, {Luppino}, {Lupton}, {Monet}, {Morgan}, {Onaka},
  {Stubbs}, {Tonry}, {Banados}, {Bell}, {Bender}, {Bernard}, {Botticella},
  {Casertano}, {Chastel}, {Chen}, {Chen}, {Cole}, {Deacon}, {Frenk},
  {Fitzsimmons}, {Gezari}, {Goessl}, {Goggia}, {Goldman}, {Grebel}, {Hambly},
  {Hasinger}, {Heavens}, {Heckman}, {Henderson}, {Henning}, {Holman}, {Hopp},
  {Ip}, {Isani}, {Keyes}, {Koekemoer}, {Kotak}, {Long}, {Lucey}, {Liu},
  {Martin}, {McLean}, {Morganson}, {Murphy}, {Nieto-Santisteban}, {Norberg},
  {Peacock}, {Pier}, {Postman}, {Primak}, {Rae}, {Rest}, {Riess}, {Riffeser},
  {Rix}, {Roser}, {Schilbach}, {Schultz}, {Scolnic}, {Szalay}, {Seitz},
  {Shiao}, {Small}, {Smith}, {Soderblom}, {Taylor}, {Thakar}, {Thiel},
  {Thilker}, {Urata}, {Valenti}, {Walter}, {Watters}, {Werner}, {White},
  {Wood-Vasey}, \& {Wyse}}]{Chambers2016}
{Chambers}, K.~C., {Magnier}, E.~A., {Metcalfe}, N., {et~al.} 2016, ArXiv
  e-prints.
\newblock \doarXiv{1612.05560}

\bibitem[{{Cloutis} {et~al.}(1990){Cloutis}, {Gaffey}, {Smith}, \&
  {Lambert}}]{Cloutis1990}
{Cloutis}, E.~A., {Gaffey}, M.~J., {Smith}, D.~G.~W., \& {Lambert}, R. S.~J.
  1990, \jgr, 95, 281, \dodoi{10.1029/JB095iB01p00281}

\bibitem[{{Cohen} {et~al.}(2023){Cohen}, {van der Bogert}, {Bottke}, {Curran},
  {Fassett}, {Hiesinger}, {Joy}, {Mazrouei}, {Nemchin}, {Neumann}, {Norman}, \&
  {Zellner}}]{Cohen2023}
{Cohen}, B.~A., {van der Bogert}, C.~H., {Bottke}, W.~F., {et~al.} 2023,
  Reviews in Mineralogy and Geochemistry, 89, 373,
  \dodoi{10.2138/rmg.2023.89.09}

\bibitem[{{Delbo} {et~al.}(2003){Delbo}, {Harris}, {Binzel}, {Pravec}, \&
  {Davies}}]{Delbo2003}
{Delbo}, M., {Harris}, A.~W., {Binzel}, R.~P., {Pravec}, P., \& {Davies}, J.~K.
  2003, \icarus, 166, 116, \dodoi{10.1016/j.icarus.2003.07.002}

\bibitem[{{Delbo} {et~al.}(2017){Delbo}, {Walsh}, {Bolin}, {Avdellidou}, \&
  {Morbidelli}}]{Delbo2017}
{Delbo}, M., {Walsh}, K., {Bolin}, B., {Avdellidou}, C., \& {Morbidelli}, A.
  2017, Science, 357, 1026, \dodoi{10.1126/science.aam6036}

\bibitem[{{DeMeo} {et~al.}(2015){DeMeo}, {Alexander}, {Walsh}, {Chapman}, \&
  {Binzel}}]{DeMeo2015}
{DeMeo}, F.~E., {Alexander}, C.~M.~O., {Walsh}, K.~J., {Chapman}, C.~R., \&
  {Binzel}, R.~P. 2015, in Asteroids IV, ed. P.~{Michel}, F.~E. {DeMeo}, \&
  W.~F. {Bottke}, 13--41

\bibitem[{{DeMeo} {et~al.}(2009){DeMeo}, {Binzel}, {Slivan}, \&
  {Bus}}]{DeMeo2009}
{DeMeo}, F.~E., {Binzel}, R.~P., {Slivan}, S.~M., \& {Bus}, S.~J. 2009,
  \icarus, 202, 160, \dodoi{10.1016/j.icarus.2009.02.005}

\bibitem[{{DeMeo} \& {Carry}(2013)}]{DeMeo2013aa}
{DeMeo}, F.~E., \& {Carry}, B. 2013, \icarus, 226, 723,
  \dodoi{10.1016/j.icarus.2013.06.027}

\bibitem[{{DeMeo} \& {Carry}(2014)}]{DeMeo2014a}
---. 2014, \nat, 505, 629, \dodoi{10.1038/nature12908}

\bibitem[{{Denneau} {et~al.}(2024){Denneau}, {Siverd}, {Tonry}, {Weiland},
  {Erasmus}, \& {Fitzsimmons}}]{Denneau2024YR4}
{Denneau}, L., {Siverd}, R., {Tonry}, J., {et~al.} 2024, Minor Planet
  Electronic Circulars, 2024-Y140

\bibitem[{{Durech} {et~al.}(2015){Durech}, {Carry}, {Delbo}, {Kaasalainen}, \&
  {Viikinkoski}}]{Durech2015}
{Durech}, J., {Carry}, B., {Delbo}, M., {Kaasalainen}, M., \& {Viikinkoski}, M.
  2015, {Asteroid Models from Multiple Data Sources}, ed. P.~{Michel}, F.~E.
  {DeMeo}, \& W.~F. {Bottke}, 183--202

\bibitem[{{Farinella} {et~al.}(1993){Farinella}, {Gonczi}, {Froeschle}, \&
  {Froeschle}}]{Farinella1993}
{Farinella}, P., {Gonczi}, R., {Froeschle}, C., \& {Froeschle}, C. 1993,
  \icarus, 101, 174, \dodoi{10.1006/icar.1993.1016}

\bibitem[{{Fukugita} {et~al.}(1996){Fukugita}, {Ichikawa}, {Gunn}, {Doi},
  {Shimasaku}, \& {Schneider}}]{Fukugita1996}
{Fukugita}, M., {Ichikawa}, T., {Gunn}, J.~E., {et~al.} 1996, \aj, 111, 1748,
  \dodoi{10.1086/117915}

\bibitem[{{Granvik} {et~al.}(2013){Granvik}, {Jedicke}, {Bolin}, {Chyba}, \&
  {Patterson}}]{Granvik2013}
{Granvik}, M., {Jedicke}, R., {Bolin}, B., {Chyba}, M., \& {Patterson}, G.
  2013, {Earth's Temporarily-Captured Natural Satellites - The First Step
  towards Utilization of Asteroid Resources}, ed. V.~{Badescu}, 151--167

\bibitem[{{Granvik} {et~al.}(2017){Granvik}, {Morbidelli}, {Vokrouhlick{\'y}},
  {Bottke}, {Nesvorn{\'y}}, \& {Jedicke}}]{Granvik2017}
{Granvik}, M., {Morbidelli}, A., {Vokrouhlick{\'y}}, D., {et~al.} 2017, \aap,
  598, A52, \dodoi{10.1051/0004-6361/201629252}

\bibitem[{{Granvik} \& {Walsh}(2024)}]{Granvik2024}
{Granvik}, M., \& {Walsh}, K.~J. 2024, \apjl, 960, L9,
  \dodoi{10.3847/2041-8213/ad151b}

\bibitem[{{Granvik} {et~al.}(2018){Granvik}, {Morbidelli}, {Jedicke}, {Bolin},
  {Bottke}, {Beshore}, {Vokrouhlick{\'y}}, {Nesvorn{\'y}}, \&
  {Michel}}]{Granvik2018}
{Granvik}, M., {Morbidelli}, A., {Jedicke}, R., {et~al.} 2018, \icarus, 312,
  181, \dodoi{10.1016/j.icarus.2018.04.018}

\bibitem[{{Hanu{\v s}} {et~al.}(2011){Hanu{\v s}}, {{\v D}urech}, {Bro{\v z}},
  {Warner}, {Pilcher}, {Stephens}, {Oey}, {Bernasconi}, {Casulli}, {Behrend},
  {Polishook}, {Henych}, {Lehk{\'y}}, {Yoshida}, \& {Ito}}]{Hanus2011}
{Hanu{\v s}}, J., {{\v D}urech}, J., {Bro{\v z}}, M., {et~al.} 2011, \aap, 530,
  A134, \dodoi{10.1051/0004-6361/201116738}

\bibitem[{{Hanu{\v s}} {et~al.}(2018){Hanu{\v s}}, {Delbo\textquotesingle},
  {Al{\'{\i}}-Lagoa}, {Bolin}, {Jedicke}, {{\v D}urech}, {Cibulkov{\'a}},
  {Pravec}, {Ku{\v s}nir{\'a}k}, {Behrend}, {Marchis}, {Antonini}, {Arnold},
  {Audejean}, {Bachschmidt}, {Bernasconi}, {Brunetto}, {Casulli}, {Dymock},
  {Esseiva}, {Esteban}, {Gerteis}, {de Groot}, {Gully}, {Hamanowa}, {Hamanowa},
  {Krafft}, {Lehk{\'y}}, {Manzini}, {Michelet}, {Morelle}, {Oey}, {Pilcher},
  {Reignier}, {Roy}, {Salom}, \& {Warner}}]{Hanus2018}
{Hanu{\v s}}, J., {Delbo\textquotesingle}, M., {Al{\'{\i}}-Lagoa}, V., {et~al.}
  2018, \icarus, 299, 84, \dodoi{10.1016/j.icarus.2017.07.007}

\bibitem[{{Hanu{\v{s}}} {et~al.}(2023){Hanu{\v{s}}}, {Vokrouhlick{\'y}},
  {Nesvorn{\'y}}, {{\v{D}}urech}, {Stephens}, {Benishek}, {Oey}, \&
  {Pokorn{\'y}}}]{Hanus2023}
{Hanu{\v{s}}}, J., {Vokrouhlick{\'y}}, D., {Nesvorn{\'y}}, D., {et~al.} 2023,
  \aap, 679, A56, \dodoi{10.1051/0004-6361/202346022}

\bibitem[{{Hanu{\v{s}}} {et~al.}(2018{\natexlab{a}}){Hanu{\v{s}}},
  {Vokrouhlick{\'y}}, {Delbo'}, {Farnocchia}, {Polishook}, {Pravec}, {Hornoch},
  {Ku{\v{c}}{\'a}kov{\'a}}, {Ku{\v{s}}nir{\'a}k}, {Stephens}, \&
  {Warner}}]{Hanus20183200}
{Hanu{\v{s}}}, J., {Vokrouhlick{\'y}}, D., {Delbo'}, M., {et~al.}
  2018{\natexlab{a}}, \aap, 620, L8, \dodoi{10.1051/0004-6361/201834228}

\bibitem[{{Hanu{\v{s}}} {et~al.}(2018{\natexlab{b}}){Hanu{\v{s}}}, {Delbo'},
  {Al{\'\i}-Lagoa}, {Bolin}, {Jedicke}, {{\v{D}}urech}, {Cibulkov{\'a}},
  {Pravec}, {Ku{\v{s}}nir{\'a}k}, {Behrend}, {Marchis}, {Antonini}, {Arnold},
  {Audejean}, {Bachschmidt}, {Bernasconi}, {Brunetto}, {Casulli}, {Dymock},
  {Esseiva}, {Esteban}, {Gerteis}, {de Groot}, {Gully}, {Hamanowa}, {Hamanowa},
  {Krafft}, {Lehk{\'y}}, {Manzini}, {Michelet}, {Morelle}, {Oey}, {Pilcher},
  {Reignier}, {Roy}, {Salom}, \& {Warner}}]{Hanus2018SpinVectorEos}
{Hanu{\v{s}}}, J., {Delbo'}, M., {Al{\'\i}-Lagoa}, V., {et~al.}
  2018{\natexlab{b}}, \icarus, 299, 84, \dodoi{10.1016/j.icarus.2017.07.007}

\bibitem[{{Harris} \& {Lagerros}(2002)}]{Harris2002}
{Harris}, A.~W., \& {Lagerros}, J.~S.~V. 2002, Asteroids III, 205

\bibitem[{{Hodgkin} {et~al.}(2009){Hodgkin}, {Irwin}, {Hewett}, \&
  {Warren}}]{Hodgkin2009}
{Hodgkin}, S.~T., {Irwin}, M.~J., {Hewett}, P.~C., \& {Warren}, S.~J. 2009,
  \mnras, 394, 675, \dodoi{10.1111/j.1365-2966.2008.14387.x}

\bibitem[{{Hook} {et~al.}(2004){Hook}, {J{\o}rgensen}, {Allington-Smith},
  {Davies}, {Metcalfe}, {Murowinski}, \& {Crampton}}]{Hook2004}
{Hook}, I.~M., {J{\o}rgensen}, I., {Allington-Smith}, J.~R., {et~al.} 2004,
  \pasp, 116, 425, \dodoi{10.1086/383624}

\bibitem[{{Ivezi{\'c}} {et~al.}(2001){Ivezi{\'c}}, {Tabachnik}, {Rafikov},
  {Lupton}, {Quinn}, {Hammergren}, {Eyer}, {Chu}, {Armstrong}, {Fan},
  {Finlator}, {Geballe}, {Gunn}, {Hennessy}, {Knapp}, {Leggett}, {Munn},
  {Pier}, {Rockosi}, {Schneider}, {Strauss}, {Yanny}, {Brinkmann}, {Csabai},
  {Hindsley}, {Kent}, {Lamb}, {Margon}, {McKay}, {Smith}, {Waddel}, {York}, \&
  {SDSS Collaboration}}]{Ivezic2001}
{Ivezi{\'c}}, {\v Z}., {Tabachnik}, S., {Rafikov}, R., {et~al.} 2001, \aj, 122,
  2749, \dodoi{10.1086/323452}

\bibitem[{{Ivezi{\'c}} {et~al.}(2002){Ivezi{\'c}}, {Lupton}, {Juri{\'c}},
  {Tabachnik}, {Quinn}, {Gunn}, {Knapp}, {Rockosi}, \&
  {Brinkmann}}]{Ivezic2002}
{Ivezi{\'c}}, {\v Z}., {Lupton}, R.~H., {Juri{\'c}}, M., {et~al.} 2002, \aj,
  124, 2943, \dodoi{10.1086/344077}

\bibitem[{{Jedicke} {et~al.}(2022){Jedicke}, {Hermosin}, {Sercel}, {Centuori},
  {Sciarra}, {Cano}, \& {Peterson}}]{Jedicke2022}
{Jedicke}, R., {Hermosin}, P., {Sercel}, J., {et~al.} 2022, \planss, 211,
  105407, \dodoi{10.1016/j.pss.2021.105407}

\bibitem[{{Jordi} {et~al.}(2006){Jordi}, {Grebel}, \& {Ammon}}]{Jordi2006}
{Jordi}, K., {Grebel}, E.~K., \& {Ammon}, K. 2006, \aap, 460, 339,
  \dodoi{10.1051/0004-6361:20066082}

\bibitem[{{Juri{\'c}} {et~al.}(2002){Juri{\'c}}, {Ivezi{\'c}}, {Lupton},
  {Quinn}, {Tabachnik}, {Fan}, {Gunn}, {Hennessy}, {Knapp}, {Munn}, {Pier},
  {Rockosi}, {Schneider}, {Brinkmann}, {Csabai}, \& {Fukugita}}]{Juric2002}
{Juri{\'c}}, M., {Ivezi{\'c}}, {\v Z}., {Lupton}, R.~H., {et~al.} 2002, \aj,
  124, 1776, \dodoi{10.1086/341950}

\bibitem[{{Kaasalainen} \& {Torppa}(2001)}]{Kaasalainen2001a}
{Kaasalainen}, M., \& {Torppa}, J. 2001, \icarus, 153, 24,
  \dodoi{10.1006/icar.2001.6673}

\bibitem[{{Kaasalainen} {et~al.}(2001){Kaasalainen}, {Torppa}, \&
  {Muinonen}}]{Kaasalainen2001b}
{Kaasalainen}, M., {Torppa}, J., \& {Muinonen}, K. 2001, \icarus, 153, 37,
  \dodoi{10.1006/icar.2001.6674}

\bibitem[{{Labrie} {et~al.}(2023){Labrie}, {Simpson}, {Cardenes}, {Turner},
  {Soraisam}, {Quint}, {Oberdorf}, {Placco}, {Berke}, {Smirnova}, {Conseil},
  {Vacca}, \& {Thomas-Osip}}]{Labrie2023}
{Labrie}, K., {Simpson}, C., {Cardenes}, R., {et~al.} 2023, Research Notes of
  the American Astronomical Society, 7, 214, \dodoi{10.3847/2515-5172/ad0044}

\bibitem[{{Larson} {et~al.}(2003){Larson}, {Beshore}, {Hill}, {Christensen},
  {McLean}, {Kolar}, {McNaught}, \& {Garradd}}]{Larson2003}
{Larson}, S., {Beshore}, E., {Hill}, R., {et~al.} 2003, in AAS/Division for
  Planetary Sciences Meeting Abstracts, Vol.~35, AAS/Division for Planetary
  Sciences Meeting Abstracts \#35, 36.04

\bibitem[{{Larson} {et~al.}(1998){Larson}, {Brownlee}, {Hergenrother}, \&
  {Spahr}}]{Larson1998}
{Larson}, S., {Brownlee}, J., {Hergenrother}, C., \& {Spahr}, T. 1998, in ,
  1037

\bibitem[{{Lomb}(1976)}]{Lomb1976}
{Lomb}, N.~R. 1976, \apss, 39, 447, \dodoi{10.1007/BF00648343}

\bibitem[{{Masiero} {et~al.}(2024){Masiero}, {Kwon}, {Dahlen}, {Masci}, \&
  {Mainzer}}]{Masiero2024}
{Masiero}, J.~R., {Kwon}, Y.~G., {Dahlen}, D.~W., {Masci}, F.~J., \& {Mainzer},
  A.~K. 2024, \psj, 5, 113, \dodoi{10.3847/PSJ/ad42a2}

\bibitem[{{Mazrouei} {et~al.}(2019){Mazrouei}, {Ghent}, {Bottke}, {Parker}, \&
  {Gernon}}]{Mazrouei2019}
{Mazrouei}, S., {Ghent}, R.~R., {Bottke}, W.~F., {Parker}, A.~H., \& {Gernon},
  T.~M. 2019, Science, 363, 253, \dodoi{10.1126/science.aar4058}

\bibitem[{{McLean} {et~al.}(2012){McLean}, {Steidel}, {Epps}, {Konidaris},
  {Matthews}, {Adkins}, {Aliado}, {Brims}, {Canfield}, {Cromer}, {Fucik},
  {Kulas}, {Mace}, {Magnone}, {Rodriguez}, {Rudie}, {Trainor}, {Wang}, {Weber},
  \& {Weiss}}]{McLean2012}
{McLean}, I.~S., {Steidel}, C.~C., {Epps}, H.~W., {et~al.} 2012, 8446, 84460J,
  \dodoi{10.1117/12.924794}

\bibitem[{{Morbidelli} {et~al.}(2020){Morbidelli}, {Delbo}, {Granvik},
  {Bottke}, {Jedicke}, {Bolin}, {Michel}, \&
  {Vokrouhlicky}}]{Morbidelli2020albedo}
{Morbidelli}, A., {Delbo}, M., {Granvik}, M., {et~al.} 2020, \icarus, 340,
  113631, \dodoi{10.1016/j.icarus.2020.113631}

\bibitem[{{Nesvorn{\'y}} {et~al.}(2015){Nesvorn{\'y}}, {Bro{\v z}}, \&
  {Carruba}}]{Nesvorny2015a}
{Nesvorn{\'y}}, D., {Bro{\v z}}, M., \& {Carruba}, V. 2015, Asteroids IV, 297,
  \dodoi{10.2458/azu_uapress_9780816530595-ch016}

\bibitem[{{Nesvorn{\'y}} {et~al.}(2023){Nesvorn{\'y}}, {Deienno}, {Bottke},
  {Jedicke}, {Naidu}, {Chesley}, {Chodas}, {Granvik}, {Vokrouhlick{\'y}},
  {Bro{\v{z}}}, {Morbidelli}, {Christensen}, {Shelly}, \&
  {Bolin}}]{Nesvorny2023NEO}
{Nesvorn{\'y}}, D., {Deienno}, R., {Bottke}, W.~F., {et~al.} 2023, \aj, 166,
  55, \dodoi{10.3847/1538-3881/ace040}

\bibitem[{{Nesvorn{\'y}} {et~al.}(2024){Nesvorn{\'y}}, {Vokrouhlick{\'y}},
  {Shelly}, {Deienno}, {Bottke}, {Fuls}, {Jedicke}, {Naidu}, {Chesley},
  {Chodas}, {Farnocchia}, \& {Delbo}}]{Nesvorny2024NEOMOD3}
{Nesvorn{\'y}}, D., {Vokrouhlick{\'y}}, D., {Shelly}, F., {et~al.} 2024,
  \icarus, 417, 116110, \dodoi{10.1016/j.icarus.2024.116110}

\bibitem[{{Popescu} {et~al.}(2018){Popescu}, {Licandro}, {Carvano},
  {Stoicescu}, {de Le{\'o}n}, {Morate}, {Boac{\u{a}}}, \&
  {Cristescu}}]{Popescu2018MOVIS}
{Popescu}, M., {Licandro}, J., {Carvano}, J.~M., {et~al.} 2018, \aap, 617, A12,
  \dodoi{10.1051/0004-6361/201833023}

\bibitem[{{Pravec} {et~al.}(2012){Pravec}, {Harris}, {Ku{\v s}nir{\'a}k},
  {Gal{\'a}d}, \& {Hornoch}}]{Pravec2012}
{Pravec}, P., {Harris}, A.~W., {Ku{\v s}nir{\'a}k}, P., {Gal{\'a}d}, A., \&
  {Hornoch}, K. 2012, \icarus, 221, 365, \dodoi{10.1016/j.icarus.2012.07.026}

\bibitem[{{Purdum} {et~al.}(2021){Purdum}, {Lin}, {Bolin}, {Sharma}, {Choi},
  {Bhalerao}, {Hanu{\v{s}}}, {Kumar}, {Quimby}, {van Roestel}, {Zhai},
  {Fernandez}, {Lisse}, {Bodewits}, {Fremling}, {Ryan Golovich}, {Hsu}, {Ip},
  {Ngeow}, {Saini}, {Shao}, {Yao}, {Ahumada}, {Anand}, {Andreoni}, {Burdge},
  {Burruss}, {Chang}, {Copperwheat}, {Coughlin}, {De}, {Dekany}, {Delacroix},
  {Drake}, {Duev}, {Graham}, {Hale}, {Kool}, {Kasliwal}, {Kostadinova},
  {Kulkarni}, {Laher}, {Mahabal}, {Masci}, {Mr{\'o}z}, {Neill}, {Riddle},
  {Rodriguez}, {Smith}, {Walters}, {Yan}, \& {Zolkower}}]{Purdum2021}
{Purdum}, J.~N., {Lin}, Z.-Y., {Bolin}, B.~T., {et~al.} 2021, \apjl, 911, L35,
  \dodoi{10.3847/2041-8213/abf2ca}

\bibitem[{{Scargle}(1982)}]{Scargle1982}
{Scargle}, J.~D. 1982, \apj, 263, 835, \dodoi{10.1086/160554}

\bibitem[{{Schwamb} {et~al.}(2023){Schwamb}, {Jones}, {Yoachim}, {Volk},
  {Dorsey}, {Opitom}, {Greenstreet}, {Lister}, {Snodgrass}, {Bolin}, {Inno},
  {Bannister}, {Eggl}, {Solontoi}, {Kelley}, {Juri{\'c}}, {Lin}, {Ragozzine},
  {Bernardinelli}, {Chesley}, {Daylan}, {{\v{D}}urech}, {Fraser}, {Granvik},
  {Knight}, {Lisse}, {Malhotra}, {Oldroyd}, {Thirouin}, \&
  {Ye}}]{Schwamb2023ApJS}
{Schwamb}, M.~E., {Jones}, R.~L., {Yoachim}, P., {et~al.} 2023, \apjs, 266, 22,
  \dodoi{10.3847/1538-4365/acc173}

\bibitem[{{Stellingwerf}(1978)}]{Stellingwerf1978}
{Stellingwerf}, R.~F. 1978, \apj, 224, 953, \dodoi{10.1086/156444}

\bibitem[{{Tonry} {et~al.}(2012){Tonry}, {Stubbs}, {Lykke}, {Doherty},
  {Shivvers}, {Burgett}, {Chambers}, {Hodapp}, {Kaiser}, {Kudritzki},
  {Magnier}, {Morgan}, {Price}, \& {Wainscoat}}]{Tonry2012}
{Tonry}, J.~L., {Stubbs}, C.~W., {Lykke}, K.~R., {et~al.} 2012, \apj, 750, 99,
  \dodoi{10.1088/0004-637X/750/2/99}

\bibitem[{{Tonry} {et~al.}(2018){Tonry}, {Denneau}, {Heinze}, {Stalder},
  {Smith}, {Smartt}, {Stubbs}, {Weiland}, \& {Rest}}]{Tonry2018}
{Tonry}, J.~L., {Denneau}, L., {Heinze}, A.~N., {et~al.} 2018, \pasp, 130,
  064505, \dodoi{10.1088/1538-3873/aabadf}

\bibitem[{{Vere{\v s}} {et~al.}(2015){Vere{\v s}}, {Jedicke}, {Fitzsimmons},
  {Denneau}, {Granvik}, {Bolin}, {Chastel}, {Wainscoat}, {Burgett}, {Chambers},
  {Flewelling}, {Kaiser}, {Magnier}, {Morgan}, {Price}, {Tonry}, \&
  {Waters}}]{Veres2015}
{Vere{\v s}}, P., {Jedicke}, R., {Fitzsimmons}, A., {et~al.} 2015, \icarus,
  261, 34, \dodoi{10.1016/j.icarus.2015.08.007}

\bibitem[{{Vernazza} {et~al.}(2009){Vernazza}, {Binzel}, {Rossi},
  {Fulchignoni}, \& {Birlan}}]{Vernazza2009}
{Vernazza}, P., {Binzel}, R.~P., {Rossi}, A., {Fulchignoni}, M., \& {Birlan},
  M. 2009, \nat, 458, 993, \dodoi{10.1038/nature07956}

\bibitem[{{Vokrouhlick{\'y}} {et~al.}(2015){Vokrouhlick{\'y}}, {Bottke},
  {Chesley}, {Scheeres}, \& {Statler}}]{Vokrouhlicky2015}
{Vokrouhlick{\'y}}, D., {Bottke}, W.~F., {Chesley}, S.~R., {Scheeres}, D.~J.,
  \& {Statler}, T.~S. 2015, Asteroids IV, 509,
  \dodoi{10.2458/azu_uapress_9780816530595-ch027}

\bibitem[{{Warner} {et~al.}(2009){Warner}, {Harris}, \& {Pravec}}]{Warner2009}
{Warner}, B.~D., {Harris}, A.~W., \& {Pravec}, P. 2009, \icarus, 202, 134,
  \dodoi{10.1016/j.icarus.2009.02.003}

\bibitem[{{Whidden} {et~al.}(2019){Whidden}, {Bryce Kalmbach}, {Connolly},
  {Jones}, {Smotherman}, {Bektesevic}, {Slater}, {Becker}, {Ivezi{\'c}},
  {Juri{\'c}}, {Bolin}, {Moeyens}, {F{\"o}rster}, \& {Golkhou}}]{Whidden2019}
{Whidden}, P.~J., {Bryce Kalmbach}, J., {Connolly}, A.~J., {et~al.} 2019, \aj,
  157, 119, \dodoi{10.3847/1538-3881/aafd2d}

\bibitem[{{Williams}(2025{\natexlab{a}})}]{Williams2025OrbitUpdate}
{Williams}, G.~V. 2025{\natexlab{a}}, Minor Planet Electronic Circulars,
  2025-D145

\bibitem[{{Williams}(2025{\natexlab{b}})}]{Williams2025MPS}
---. 2025{\natexlab{b}}, Minor Planet Electronic Circulars Supplement, 2321551

\bibitem[{{Willmer}(2018)}]{Willmer2018}
{Willmer}, C. N.~A. 2018, \apjs, 236, 47, \dodoi{10.3847/1538-4365/aabfdf}

\bibitem[{{Wisdom}(1983)}]{Wisdom1983}
{Wisdom}, J. 1983, \icarus, 56, 51, \dodoi{10.1016/0019-1035(83)90127-6}

\end{thebibliography}

\clearpage
\newpage
\renewcommand{\thefigure}{A\arabic{figure}}
\setcounter{figure}{0}
\renewcommand{\thetable}{A\arabic{table}}
\renewcommand{\theequation}{A\arabic{equation}}
\renewcommand{\thesection}{A\arabic{section}}
\setcounter{section}{0}
\clearpage
\setcounter{page}{1}
\renewcommand\thepage{A\arabic{page}}
\appendix
\renewcommand{\thefigure}{A\arabic{figure}}
\setcounter{figure}{0}
\renewcommand{\thetable}{A\arabic{table}}
\renewcommand{\theequation}{A\arabic{equation}}
\renewcommand{\thesection}{A}
\setcounter{section}{0}
\setcounter{table}{0}
\begin{longtable}{|c|c|c|c|}
\caption{Summary of \yr photometry from between 2024 December 27 and 2025 February 7 UTC.\label{t.photometry1}}\\
\hline
Date$^1$ & Mag$^2$ & Mag unc.$^3$&Obs. Code$^4$ \\
(MJD UTC)&&(s)&\\
\hline
\endfirsthead
\multicolumn{4}{c}%
{\tablename\ \thetable\ -- \textit{Continued from previous page}} \\
\hline
Date$^1$ & Mag$^2$ & Mag unc.$^3$&Obs. Code$^4$ \\
(MJD UTC)&&(s)&\\
\hline
\endhead
\hline \multicolumn{4}{r}{\textit{Continued on next page}} \\
\endfoot
\hline
\endlastfoot
60671.23807315 & 16.54 & 0.03 & W68 \\
60671.24033426 & 15.80 & 0.03 & W68 \\
60671.26732187 & 16.28 & 0.03 & W68 \\
60671.27464815 & 16.12 & 0.03 & W68 \\
60672.34588600 & 17.31 & 0.10 & 703 \\
60672.35043700 & 17.12 & 0.10 & 703 \\
60672.35271000 & 17.26 & 0.10 & 703 \\
60672.48815185 & 17.10 & 0.05 & T05 \\
60672.48996875 & 17.30 & 0.05 & T05 \\
60672.49358611 & 17.14 & 0.03 & T05 \\
60672.49902662 & 16.98 & 0.03 & T05 \\
60673.52814200 & 17.49 & 0.13 & F52 \\
60673.54101000 & 17.13 & 0.13 & F52 \\
60673.55386300 & 17.35 & 0.13 & F52 \\
60673.56663600 & 17.37 & 0.13 & F52 \\
60674.40934711 & 18.05 & 0.07 & T08 \\
60674.41258137 & 18.07 & 0.07 & T08 \\
60674.41536956 & 18.00 & 0.05 & T08 \\
60674.44491667 & 17.72 & 0.04 & T08 \\
60675.28366667 & 18.04 & 0.09 & W68 \\
60675.28781644 & 18.45 & 0.09 & W68 \\
60675.30947500 & 18.27 & 0.27 & 703 \\
60675.31180000 & 18.38 & 0.27 & 703 \\
60675.31644000 & 18.90 & 0.27 & 703 \\
60675.32319479 & 18.04 & 0.07 & W68 \\
60675.34909931 & 18.11 & 0.12 & W68 \\
60676.55924286 & 18.80 & 0.11 & T05 \\
60676.56982976 & 18.68 & 0.09 & T05 \\
60676.57969515 & 18.55 & 0.11 & T05 \\
60677.43338700 & 18.72 & 0.11 & F52 \\
60677.44491300 & 18.82 & 0.11 & F52 \\
60677.45652500 & 19.02 & 0.11 & F52 \\
60677.46819900 & 18.87 & 0.11 & F52 \\
60678.32803200 & 19.25 & 0.18 & 703 \\
60678.33011000 & 19.60 & 0.18 & 703 \\
60678.33218800 & 19.29 & 0.18 & 703 \\
60678.33426500 & 19.09 & 0.18 & 703 \\
60678.45297894 & 19.23 & 0.15 & T08 \\
60678.45618218 & 19.24 & 0.15 & T08 \\
60678.45939005 & 19.00 & 0.13 & T08 \\
60678.48978009 & 19.00 & 0.12 & T08 \\
60680.33804700 & 19.22 & 0.17 & 703 \\
60680.34013000 & 19.37 & 0.17 & 703 \\
60680.34428800 & 19.63 & 0.17 & 703 \\
60680.39489045 & 19.24 & 0.17 & T05 \\
60680.44102885 & 19.38 & 0.19 & T05 \\
60696.11607700 & 21.55 & 0.05 & 309 \\
60696.11669800 & 21.54 & 0.05 & 309 \\
60696.11851100 & 21.55 & 0.05 & 309 \\
60696.11911300 & 21.42 & 0.04 & 309 \\
60696.12032800 & 21.43 & 0.04 & 309 \\
60696.12093800 & 21.52 & 0.05 & 309 \\
60696.12155200 & 21.53 & 0.05 & 309 \\
60696.12215700 & 21.57 & 0.05 & 309 \\
60696.12275900 & 21.57 & 0.05 & 309 \\
60696.12337200 & 21.55 & 0.05 & 309 \\
60696.12398200 & 21.42 & 0.04 & 309 \\
60696.12458800 & 21.34 & 0.04 & 309 \\
60696.12520900 & 21.34 & 0.04 & 309 \\
60696.12581400 & 21.22 & 0.04 & 309 \\
60696.12642500 & 21.15 & 0.03 & 309 \\
60696.12703500 & 21.23 & 0.04 & 309 \\
60696.12764300 & 21.24 & 0.04 & 309 \\
60696.12826200 & 21.22 & 0.04 & 309 \\
60696.12887900 & 21.39 & 0.04 & 309 \\
60696.12948400 & 21.53 & 0.05 & 309 \\
60696.13008500 & 21.65 & 0.05 & 309 \\
60696.13069600 & 21.73 & 0.05 & 309 \\
60696.13130900 & 21.69 & 0.05 & 309 \\
60696.13191400 & 21.49 & 0.04 & 309 \\
60696.13253600 & 21.46 & 0.04 & 309 \\
60696.13314100 & 21.51 & 0.04 & 309 \\
60696.13375100 & 21.51 & 0.05 & 309 \\
60696.13619300 & 21.48 & 0.04 & 309 \\
60696.13679900 & 21.51 & 0.04 & 309 \\
60696.13740900 & 21.36 & 0.04 & 309 \\
60696.13801400 & 21.32 & 0.04 & 309 \\
60696.13862800 & 21.24 & 0.04 & 309 \\
60696.13923000 & 21.18 & 0.03 & 309 \\
60696.13984300 & 21.21 & 0.04 & 309 \\
60696.14045700 & 21.18 & 0.03 & 309 \\
60696.14107800 & 21.17 & 0.03 & 309 \\
60697.33390700 & 22.01 & 0.20 & F52 \\
60697.34534000 & 21.92 & 0.20 & F52 \\
60697.35679700 & 21.51 & 0.20 & F52 \\
60697.36823700 & 21.64 & 0.20 & F52 \\
60713.10456210 & 23.38 & 0.22 & I11 \\
60713.12729625 & 23.63 & 0.29 & I11 \\
60713.13864208 & 23.38 & 0.19 & I11 \\
60713.14415887 & 23.35 & 0.19 & I11 \\
60713.15552038 & 23.51 & 0.23 & I11 \\
60713.16686808 & 23.29 & 0.17 & I11 \\
60713.19509238 & 23.48 & 0.18 & I11 \\
60713.20646260 & 23.21 & 0.14 & I11 \\
60713.10180008 & 23.24 & 0.11 & I11 \\
60713.11316511 & 23.56 & 0.14 & I11 \\
60713.14139774 & 23.44 & 0.12 & I11 \\
60713.15275886 & 23.35 & 0.12 & I11 \\
60713.15827524 & 23.11 & 0.08 & I11 \\
60713.16962233 & 23.21 & 0.09 & I11 \\
60713.18097070 & 23.42 & 0.10 & I11 \\
60713.19232942 & 23.38 & 0.10 & I11 \\
60713.10734983 & 23.63 & 0.12 & I11 \\
60713.12143592 & 23.73 & 0.14 & I11 \\
60713.13280506 & 23.38 & 0.09 & I11 \\
60713.14694628 & 23.40 & 0.09 & I11 \\
60713.17237680 & 23.33 & 0.09 & I11 \\
60713.18652172 & 23.35 & 0.09 & I11 \\
60713.20061316 & 23.32 & 0.09 & I11 \\
60713.11023326 & 23.32 & 0.22 & I11 \\
60713.13568818 & 23.52 & 0.26 & I11 \\
60713.14983347 & 23.31 & 0.22 & I11 \\
60713.16391275 & 23.39 & 0.25 & I11 \\
60713.18940483 & 23.54 & 0.26 & I11 \\
60713.20350766 & 23.37 & 0.29 & I11 \\
\hline
\caption{Columns: (1) observation date; (2) r-band equivalent magnitude; (3) 1-$\sigma$ magnitude uncertainty (4) observatory code}
\label{t:photo}
\end{longtable}
\end{document}